\documentclass[a4paper,12pt]{article}
\pdfoutput=1
\usepackage{graphicx, rotating,amssymb,amsmath,enumitem,latexsym}

\ifx\pdfoutput\undefined
\usepackage[dvips,bookmarks]{hyperref}	
\else
\usepackage{hyperref}	
\fi
\hypersetup{colorlinks,bookmarksopen,bookmarksnumbered,citecolor=rossoc,
linkcolor=verdes,pdfstartview=FitH,urlcolor=rosso}
\def\myurl#1#2{\href{http://#1}{#2}}
\def\hhref#1{\href{http://arxiv.org/abs/#1}{#1}} 

\usepackage{multicol}
\usepackage{color}
\definecolor{rosso}{cmyk}{0,1,1,0.4}
\definecolor{rossos}{cmyk}{0,1,1,0.55}
\definecolor{rossoc}{cmyk}{0,1,1,0.2}
\definecolor{blu}{cmyk}{1,1,0,0.3}
\definecolor{blus}{cmyk}{1,1,0,0.6}
\definecolor{bluc}{cmyk}{1,1,0,0.1}
\definecolor{verde}{cmyk}{0.92,0,0.59,0.25}
\definecolor{verdec}{cmyk}{0.92,0,0.59,0.15}
\definecolor{verdes}{cmyk}{0.92,0,0.59,0.4}

\definecolor{rossoCP3}{cmyk}{0,.88,.77,.40}

\begin{document}

\setcounter{page}{0}
\thispagestyle{empty}

\begin{flushright}
\small{
CERN-PH-TH/2011-045\\
SACLAY--T11/022\\
CP$^3$-Origins-2011-031\\ 
DIAS-2011-23}
\end{flushright}

\color{black}
 
\vskip 10pt

\begin{center}

{\bf \LARGE {
Consequences of DM/antiDM Oscillations\\
\vskip 8pt
for Asymmetric WIMP Dark Matter}}

\vskip 12pt

\color{black}\vspace{0.8cm}

{
{\large\bf Marco Cirelli}$^{a,b}$,
{\large\bf  Paolo Panci}$^{a,c,d,e,f}$,\\[2mm]
{\large\bf  G\'eraldine Servant}$^{a,b}$,
{\large\bf  Gabrijela Zaharijas}$^{b,g}$,
}

\vspace{0.5cm}

{\footnotesize\noindent
{\it $^a$ CERN Theory Division, CH-1211 Gen\`eve, Switzerland}\\[0.1cm]
{\it $^b$ Institut de Physique Th\'eorique, CNRS, URA 2306 \& CEA/Saclay,
	F-91191 Gif-sur-Yvette, France}\\[0.1cm]
{\it $^c$ Dipartimento di Fisica, Universit\`a degli Studi dell'Aquila, 67010 Coppito (AQ)}\\[0.1cm]
{\it $^d$ INFN, Laboratori Nazionali del Gran Sasso, 67010 Assergi (AQ), Italy}\\[0.1cm]
{\it $^e$ Universit\'e Paris 7-Diderot, UFR de Physique, 10, rue A. Domon et L.Duquet, 75205 Paris, France}\\[0.1cm]
{\it $^f$ CP3-origins \& the Danish Institute for Advanced Study { DIAS},\\ 
University of Southern Denmark, Campusvej 55, DK-5230 Odense M, Denmark}\\[0.1cm]
{\it $^g$  Institut d'Astrophysique de Paris, UMR 7095-CNRS \& \\[-0.1cm] 
Universit\'e Pierre et Marie Curie, boulevard Arago 98bis, 75014, Paris, France.}}

\end{center}

\vspace{.75cm}

\begin{abstract}

Assuming the existence of a primordial asymmetry in the dark sector, a scenario usually dubbed Asymmetric Dark Matter (aDM), we study the effect of oscillations between dark matter and its antiparticle on the re-equilibration of  the initial asymmetry before freeze-out, which enable efficient annihilations to recouple. We calculate the evolution of  the DM relic abundance and show how oscillations re-open the parameter space of aDM models, in particular in the direction of allowing large (WIMP-scale) DM masses.
A typical wimp with a mass at the EW scale ($\sim$ 100 GeV $-$ 1 TeV) presenting a primordial asymmetry of the same order as the baryon asymmetry naturally gets the correct relic abundance if the DM-number-violating  $\Delta$(DM) = 2 mass term is in the $\sim$ meV range.
The re-establishment of annihilations implies that constraints from the accumulation of aDM in astrophysical bodies are evaded. On the other hand, the ordinary bounds from BBN, CMB and indirect detection signals on annihilating DM have to be considered.

\end{abstract}

\newpage

\section{Introduction}
\label{introduction}

Dark Matter (DM) constitutes a sizable fraction of the current energy density of the Universe, corresponding to $\Omega_{\mbox{\tiny DM}} h^2 =0.1126 \pm 0.0036$~\cite{cosmoDM}\footnote{Here $\Omega_{\mbox{\tiny DM}} = \rho_{\mbox{\tiny DM}}/\rho_c$ is defined as usual as the energy density in dark matter with respect to the critical energy density of the Universe $\rho_c = 3 H_0^2/8\pi G_N$, where $H_0$ is the present Hubble parameter. $h$ is its reduced value $h = H_0 / 100\ {\rm km}\, {\rm s}^{-1} {\rm Mpc}^{-1}$.}. 
A leading explanation for DM in the last three decades has been to postulate the existence  of a cosmologically stable weakly-interacting massive particle (WIMP), as well-motivated by Standard Model extensions at the electroweak scale. Most of the proposed candidates arising in these frameworks are either their own antiparticles (e.g. the Majorana neutralino in SUSY) or it is assumed that DM particles and anti-particles are produced in equal numbers, in contrast to what happens in the visible sector, i.e. for baryons. The origin of DM in both cases is explained in terms of the standard freeze-out mechanism, by which the annihilations of DM and its anti-particle naturally stop when the expansion of the universe overcomes the strength of their cross section, thus leaving the current DM abundance as the left-over of an incomplete annihilation process. The only crucial parameter setting $\Omega_{\mbox{\tiny DM}}$, in this standard framework, is indeed the annihilation cross section.

\smallskip

With the experimental programme to search for WIMPs being close to reaching its culmination point,
theorists have increasingly considered deviations with respect to standard paradigm presented above. One route is to assume a different cosmological history \cite{Gelmini:2010zh}, or to consider very feebly interacting particles which would have never reached thermal equilibrium \cite{Boyarsky:2009ix,Hall:2009bx}. 

Another possibility is to assume that DM particles were once in thermal equilibrium {\em with an initial asymmetry} between particles and anti-particles, as originally considered in Technicolor-like constructions~\cite{Nussinov:1985xr, Barr:1990ca, Barr:1991qn, Kaplan:1991ah,GudnasonKouvarisSannino} or mirror models~\cite{mirror1,mirror2,mirror3,mirror4,mirror5,mirror6,mirror7}, but also in other contexts \cite{Farrar:2005zd,Hooper:2004dc,Kitano:2004sv,Agashe:2004bm,Cosme:2005sb,Khlopov:2005ew}. 
In the latest two years, there has been a revival of interest for this scenario, dubbed Asymmetric Dark Matter (aDM)~\cite{Kaplan:2009ag, Cohen:2009fz, Cai:2009ia,An:2009vq, Shelton:2010ta, Buckley:2010ui, Davoudiasl:2010am, Haba:2010bm, Chun:2010hz, Gu:2010ft, Blennow:2010qp, McDonald:2010rn, Allahverdi:2010rh, Dutta:2010va, Falkowski:2011xh,Cheung:2011if,DelNobile:2011je,Cui:2011qe,MarchRussell:2011fi,Frandsen:2011kt,Buckley:2011kk,Davoudiasl:2011fj,Graesser:2011vj,Arina:2011cu,McDonald:2011sv,Barr:2011cz}, with the aim in particular of connecting the DM abundance to the abundance of baryons, i.e. to understand the origin of the ratio $\Omega_{\mbox{\tiny B}}/\Omega_{\mbox{\tiny DM}}\sim 1/5$.
 ($\Omega_{\mbox{\tiny B}} h^2 = 0.0226 \pm 0.00053$~\cite{cosmoDM}). 
 A common production history for the dark and visible matter, in fact, provides an elegant explanation of why the two densities are so close to each other. This approach, in its simplest realizations, suggests a rather light particle, $\cal{O}$(5 GeV): this does not match the expected scale of new physics, but part of the community has seen in it intriguing connections with some recent hints of signals in various direct detection experiments~\cite{direct}.\footnote{Some works have however shown how the connection between the baryon asymmetry and the DM asymmetry can be preserved even for DM particles with masses in the WIMP ($100 \mbox{ GeV}-1$ TeV) range e.g.~\cite{Buckley:2010ui}.}
Like for the baryonic abundance, if there is an asymmetry in the dark sector, as soon as annihilations have wiped out the density of (say) antiparticles, the number density of particles remains frozen for lack of targets, and is entirely controlled by the primordial asymmetry rather than by the value of the annihilation cross section.  This is why this scenario appears rather constraining on the value of the DM mass. 

\medskip

This conclusion, however, changes in the presence of oscillations between DM and antiDM particles, and it is the purpose of this paper to study this in detail. Such oscillations can indeed replenish the depleted population of `targets'. Annihilations, if strong enough, can then re-couple and deplete further the DM/antiDM abundance. The final DM relic abundance is therefore attained through a more complex history than in the standard case of aDM, and in closer similarity to the freeze-out one.
Somehow, the phenomenology associated to oscillations has not been yet studied in detail.
The effect of oscillations between DM and $\overline{{\rm DM}}$ was  mentioned only a few times in the literature~\cite{Cohen:2009fz,Cai:2009ia, Chun:2010hz, Falkowski:2011xh,Arina:2011cu}. 
In these works, an oscillation mechanism between particles and antiparticles was considered only at late times. In fact, oscillations were prevented by assumption from occurring too early, when annihilations are still coupled, in order to maintain the simple relation between the DM asymmetry and the baryon asymmetry. 
We are in contrast interested in the opposite situation where oscillations do play a role and control the final relic abundance. This situation is possible and not unlikely: there is no specific physical reason why oscillations could not start early on, and therefore  this situation should not be disregarded. Moreover, it is an instructive setup in the sense that it fills a gap between the standard thermal freeze out prediction (where $\Omega_{\mbox{\tiny DM}}$ does not depend explicitely on the DM mass but only on the annihilation cross section $\langle \sigma v \rangle$), and the aDM prediction where $\Omega_{\mbox{\tiny DM}} h^2$ does not depend on $\langle \sigma v \rangle$ but only on the primordial DM asymmetry.

\medskip

Another interesting consequence of adding oscillations on top of aDM concerns the phenomenological bounds. Bounds on the traditional aDM framework have been studied in several works~\cite{bounds}: most constraints follow from possible decays of aDM particles or from the effect of accumulation in stars. If, however, aDM annihilates again at late times, as it does in the scenarios that we are considering, most of such bounds are evaded in a natural way. 
On the other hand, the revival of annihilations leads to indirect detection signals and subjects our framework to the usual constraints on annihilating DM, as we will discuss below.

\medskip

The rest of the paper is organized as follows. In Section~\ref{sec:theory} we briefly review some theory motivations for having DM/antiDM oscillations and we make contact between the phenomenological parameter $\delta m$, which enters in the physics of oscillations, and the scales of possible underlying particle physics models. In Sec.~\ref{sec:formalism} we lay down the formalism that we use for treating the system of annihilating and oscillating DM particles, and we illustrate the outcome in a few illustrative cases. We also include in the treatment the elastic scatterings that DM particles have with the primordial plasma and see that these can have an important effect in modifying the evolution.
In~\cite{Graesser:2011wi} and~\cite{Iminniyaz:2011yp} a detailed study of the evolution (Boltzmann) equations of the populations of aDM has been performed: our work is  a generalization of these results to the case in which DM and $\overline{{\rm DM}}$ oscillations also happen. In Sec.~\ref{sec:results} we present more systematically the results as a function of the choices of parameters in the system, and individuate the interesting regions of the parameter space. In Sec.~\ref{sec:constraints} we discuss the impact of the current constraints (from cosmology, astrophysics and colliders) on our parameter space. 
Finally, Sec.~\ref{conclusions} summarizes our conclusions.


\section{Theory motivations}
\label{sec:theory}

In this work we assume that the dark matter particle DM is not its antiparticle $\overline{{\rm DM}}$ and we assume that there is a primordial asymmetry between the two populations. 
We will follow a phenomenological approach, in the sense that we are agnostic about the the origin of such primordial asymmetry: we only assume its existence and we study 
the evolution of the two populations  in the presence of oscillations
generated by a $\Delta ({\rm DM})=2$ mass  term,  $\delta m$.
 We assume that all operators responsible for the asymmetry
are switched off when we start following the evolution, which is reasonable
 when considering WIMP-scale particles.

The effect of $\Delta ({\rm DM})=2$ operators is to introduce a mass splitting and mixing between DM and $\overline{{\rm DM}}$, which are no longer mass eigenstates.
 Oscillations will be cosmologically relevant if $\delta m \gtrsim H \sim T^2/m_{Pl}$. Therefore, it is clear that for a too large $\delta m$, oscillations will start too early, well before annihilations freeze-out and we recover a standard symmetric DM freeze-out scenario. If on the other hand, $\delta m$ is small, oscillations may start during or after annihilations freeze-out, leading to an interesting new phenomenology modifying the final DM relic abundance, a situation that has not yet been studied in detail.

Let us first consider the case where DM is a fermion and both Majorana and Dirac masses are present. The general mass lagrangian using the Weyl spinors $X_L$ and $X_R$ is given by 
\begin{eqnarray}
- {\cal L}_{mass}= m \ (\overline{X_R} X_L + \overline{X_L} X_R) + \ \Delta \ ( \overline{X_L} (X_L)^c + \overline{(X_R)^c} X_R)
\end{eqnarray}
where we have assumed $\Delta_L=\Delta_R=\Delta$ for simplicity.
In  matricial form it becomes
\begin{eqnarray}
-{\cal L}_{mass}=\frac{1}{2} \   \overline{ \left((X_L)^c \ \ X_R \right) }
\left(
\begin{array}{cc}
\Delta  & m \\
 m  & \Delta   
\end{array}
\right)
\left(
\begin{array}{c}
X_L\\
 (X_R)^c  
\end{array}
\right) + h.c.
\end{eqnarray}
The matrix
${\cal M}= \left(
\begin{array}{cc}
\Delta  & m \\
 m  & \Delta    
\end{array}
\right) $
is symmetric due to the anti commutation properties of the fermion fields  and the properties of the charge conjugation matrix $C$ \cite{Bilenky:1987ty}.
It has  mass eigenvalues $m_{1/2}=m \mp \Delta$, associated with mass eigen states $X_{1/2,L}=  (X_R)^c \mp X_L$, $X_{1/2,R}=  X_R \mp X_L^c$. We can then deduce the effective hamiltonian in the non-relativistic limit in the $(X,X^c)$ basis:
\begin{equation}
{\cal H}={\cal U}^{-1} 
\left(
\begin{array}{cc}
m-\Delta  & 0 \\
 0  & m+\Delta    
\end{array}
\right)
 {\cal U}
 =
 \left(
 \begin{array}{cc}
m&\Delta   \\
 \Delta & m    
\end{array}
\right)
\label{eq:hamiltonianmatrix}
\end{equation}
A non-zero value for $\Delta$ is  responsible for the oscillations between $X$ and $X^c$. We will typically be considering the situation $\Delta \ll m$
\footnote{Note that one can easily generalize our results to the case where $\Delta_{L} \neq \Delta_{R}$. In terms of $\Delta_+=\Delta_{L}+\Delta_R$ and $\Delta_-=\Delta_{L}-\Delta_{R}$, the mass eigenstates become
$m_{1/2}=\frac{1}{2}(\Delta_+\mp\sqrt{\Delta_-^2+4m^2})$. Since we work in the regime $\Delta_{L}, \Delta_{R} \ll 1$, one just has to replace $\Delta$  in (\ref{eq:hamiltonianmatrix}) by $\Delta=(\Delta_{L}+\Delta_{R})/2$. 
As for the mass eigenstates, $X_{1/2,L}=(\frac{\Delta_-\mp \sqrt{\Delta_-^2+4m^2}}{2m},1)$, they remain almost-equal admixtures of $X$ and $X^c$.}. 

A similar analysis applies to a complex scalar
field which splits into two quasi-degenerate real scalars. A well-known example is the sneutrino that carries the same lepton numbers as the neutrino and is distinct from its antiparticle, the anti-sneutrino. In the presence of a lepton number violation (for instance through the $\tilde{l}\tilde{l} H H$ operator, where $H$ is the Higgs field) sneutrinos can mix with anti-sneutrinos since no other quantum  numbers forbid the mixing \cite{Hirsch:1997vz,Grossman:1997is,Hall:1997ah,Choi:2001fka}.
The mass squared matrix can be written for a single generation as
\begin{eqnarray}
{\cal L}_{mass}=\frac{1}{2} \  \left( \varphi , \ \varphi^* \right)^* 
\left(
\begin{array}{cc}
m^2 &\Delta^2/2 \\
 \Delta^2/2 & m^2   
\end{array}
\right)
\left(
\begin{array}{c}
\varphi\\
 \varphi^* 
\end{array}
\right) 
\end{eqnarray}
The mass eigenvalues are now $m_{1/2}^2=m^2 \mp \Delta^2/2$ so that for $\Delta \ll m$, 
$m_2 -m_1 \approx \Delta^2/(2M)$. Therefore, in the bosonic case, the mass splitting between the mass eigenstates is given by the see-saw formula, $\Delta^2/(2M)$, rather than the mass term breaking the DM number, $2\Delta$, and this factor is what enters in the off-diagonal component of the effective lagrangian. In the $(\varphi,\varphi^*)$ basis:
\begin{equation}
{\cal H} =
 \left(
 \begin{array}{cc}
m&\Delta^2/(4M)   \\
 \Delta^2/(4M) & m    
\end{array}
\right)
\end{equation}
Therefore, for our phenomenological analysis, we will use the generic form
\begin{equation}
{\cal H} =
 \left(
 \begin{array}{cc}
m&\delta m   \\
 \delta m & m    
\end{array}
\right)
\ \ \mbox{where } \ \ 
 \delta m=
  \left\{
 \begin{array}{cc}
\Delta  &  \mbox{ if fermionic DM}\\
\Delta^2/(4M) &     \mbox{ if bosonic DM}
\end{array}
\right.
\end{equation}

Note that the lagrangians of the models we are concerned with are similar  to the ones of inelastic dark matter \cite{TuckerSmith:2001hy,Cui:2009xq}, although we are considering a much smaller  $\Delta$ so that at the end we are focussing on different  phenomenological properties. Typical examples are either  a WIMP interacting with a hidden $U(1)'$ gauge boson or a WIMP charged under  $SU(2)_L$
\cite{Cui:2009xq,Arina:2011cu}.
Both in the fermionic and bosonic cases, it is technically natural to have the `Majorana' mass $\Delta$ much smaller than the `Dirac' mass $m$ since $\Delta$ violates a global $U(1)_{\mbox{\tiny DM}}$ symmetry, due for instance to the {\it vev} of some scalar field  and all quantum corrections to $\Delta$ are proportional to itself. In our model-independent study,  $\delta m$ is a free parameter which, even if very small, will be scanned over orders of magnitude in the sub-eV range. 
Still, let us note that a natural value in the fermionic case is obtained from  the dimension-5 operator 
\begin{equation}
 \frac{XXH^{\dagger}H}{\Lambda}
\end{equation}
After electroweak symmetry breaking  and taking $\Lambda$ at the Planck scale we obtain the see-saw value $\delta m \sim 10^{-6}$ eV.
This value turns out to lead to interesting cosmological effects.
In fact,  as we will see shortly, if $m\lesssim 10 $ TeV, $\delta m$ should not be larger than $\sim 1$ eV if we want oscillations to have an effect on the final relic abundance.
In the bosonic case, this translates into  a bound $\Delta \lesssim 10^{-2}$ GeV, which is less straightforward to explain from an operator
\begin{equation}
\lambda \ {\varphi \varphi HH}
\end{equation}
since that would require $\lambda \lesssim 10^{-8}$.
There are however ways to sequester the effects of $U(1)_{\mbox{\tiny DM}}$ breaking, see 
e.g. \cite{Cui:2009xq}.

Since the upper edge of cosmologically relevant  values for $\delta m$  may not be so far away from the mass scale of neutrinos, it is tempting to try and link the two. Even when the two scales vary by orders of magnitude, it is worth considering 
a possible common origin for the Majorana masses of neutrino and dark matter. 
There is a significant literature which relates neutrino mass and dark matter (e.g. \cite{Lindner:2011it} and references therein).
There has also been attempts to link DM and neutrinos together  with leptogenesis. For instance, in the recent 
Ref. \cite{Falkowski:2011xh}, an extra hidden scalar $\phi$ couples DM with the right-handed neutrino $N$. In this class of models, if $\phi$ acquires a {\it vev}, it generates a Majorana mass for DM but also induces a mixing between DM and neutrinos that can lead to DM decay depending on the choice of parameters, in particular on $m_N$.  Alternatively,
an earlier interesting possibility was brought up in \cite{Ma:2006km} where a $\mathbb{Z}_2$ symmetry forbids a {\it vev} for the new scalar (an $SU(2)_L$ doublet) and there is no Dirac mass linking $\nu$ with $N$, thus guaranteeing the stability of DM. Nevertheless a Majorana mass can be generated at loop-level. Another explanation for the stability of DM may be that a $\mathbb{Z}_2$ emerges as an unbroken  remnant  of a global $U(1)_{B-L}$  \cite{Lindner:2011it}.


\section{Oscillation + annihilation + scattering formalism}
\label{sec:formalism}

Our aim is to study the evolution in time $t$ of the populations of DM particles and their antiparticles $\overline{{\rm DM}}$, denoted respectively by $n^+$ and $n^-$,
which possess an initial asymmetry and are subject to the simultaneous processes of annihilations ${\rm DM}\, \overline{{\rm DM}} \rightarrow {\rm SM}\, \overline{{\rm SM}}$ (with ${\rm SM}$ being any Standard Model particle), oscillations ${\rm DM} \leftrightarrow \overline{{\rm DM}}$ and elastic scatterings ${\rm DM}\, {\rm SM} \rightarrow {\rm DM}\, {\rm SM}$. 
For definiteness, we assume that particles are initially more abundant than antiparticles, i.e. $n^+ > n^-$.

The proper tool to treat this problem, in which a coherent process such as oscillations is overlapping with incoherent processes such as annihilations and scatterings, is provided by the density matrix formalism, originally developed for the case of neutrino oscillations in the Early Universe~\cite{formalism}, but which can be adapted to our present needs. One defines a $2\times 2 $ matrix, whose diagonal entries correspond to the individual number densities $n^+$ and $n^-$ and whose off-diagonal entries express the superposition of quantum states ${}^+$ and ${}^-$ originated by the oscillations.  
As is customary, we introduce the comoving densities $Y^\pm \equiv n^\pm/s$, where $s$ is the total entropy density of the Universe, and we follow the evolution in terms of the dimensionless variable $x= m_{\mbox{\tiny DM}}/T$, where $m_{\mbox{\tiny DM}}$ is the  DM mass and $T$ the temperature. We will therefore work in terms of a {\em comoving number density matrix} 
\begin{equation}
\label{densitymatrix}
\mathcal{Y}(x) = \left( \begin{array}{cc} Y^+(x) & Y^{+-}(x)\\ Y^{-+}(x) & Y^-(x) \end{array} \right)
\end{equation}
(the curly font for $\mathcal{Y}$ will indicate in the following the matrix quantity). We will always be interested in the epoch of radiation domination, during which the Hubble parameter $H(x)=\sqrt{8\pi^3 g_*(x)/90}\, m_{\mbox{\tiny DM}}^2  x^{-2}/M_{\rm Pl}=H_m/x^2$ and $t^{-1} = 2 H(x)$. In terms of $x$ one also has $s(x) \simeq 2\pi^2/45 \,g_{*\rm s}(x) \, m_{\mbox{\tiny DM}}^3 \cdot (1/x^3)$.~\footnote{The $\simeq$ sign in the latter relation just reminds that the total entropy density is  dominated by the entropy density in relativistic degrees of freedom, in a very good approximation.} Here $g_*(x)$ and $g_{*\rm s}(x)$ are the effective relativistic degrees of freedom. We define the $^\prime$ notation as 
\begin{equation}
\  ^\prime  \equiv  \left[ 1-\frac{x}{4} \frac{dg_*(x)/dx}{g_*(x)} \right]^{-1} \times  \frac{d\, }{dx} = \frac{1 }{x\, H(x)} \times \frac{d\, }{dt} 
\label{jacob}
\end{equation}
Neglecting the $x$-dependence of $g_*$  is often an acceptable approximation; for completeness, however, we keep the factor in square brackets in eq.~(\ref{jacob}) in all our computations.

We will now write explicitly the full density matrix equation that we consider. For a better illustration and understanding, we will discuss each piece of the equation (and the parameters that they contain) one by one in the next subsections, considering in turn a situation with only annihilations and no oscillations nor elastic scatterings, a situation with oscillations only, then combining oscillations and annihilations and finally including the elastic scattering as well. In the cases in which it is possible and convenient, we will deduce from the matricial form of the equation the more familiar Boltzmann equations for $Y^+$ and $Y^-$. The evolution equation for the density matrix $\mathcal{Y}$ reads
\begin{eqnarray}
\label{masterequation}
\mathcal{Y}^{\, \prime}(x) & = & -  \frac{i}{x\, H(x)} \Big[\mathcal{H},\mathcal{Y}(x) \Big] \\  \nonumber
&  & - \frac{s(x)}{x\, H(x)} \left( \frac{1}{2} \Big\{ \mathcal{Y}(x), \Gamma_{\rm a}\, \bar{\mathcal{Y}}(x) \, \Gamma_{\rm a}^\dagger \Big\}  -  \Gamma_{\rm a} \, \Gamma_{\rm a}^\dagger \, \mathcal{Y}_{\rm eq}^2 \right) \\  \nonumber
& & - \frac{1}{x\, H(x)} \Big\{ \Gamma_{\rm s}(x), \mathcal{Y}(x) \Big\}. 
\end{eqnarray}
On the right hand side, the first term accounts for oscillations, the second for annihilations and the third for elastic scatterings. 
The initial conditions read $Y^\pm_0 \equiv Y^\pm (x_0) = Y_{\rm eq}(x_0) \, e^{\pm \xi_0}$ and $Y^{+-}(x_0) = Y^{-+}(x_0) =0$, at an initial time $x_0$ (in practice we usually choose $x_0 = 5$, early enough to be able to follow the whole subsequent evolution, but not too early, so that we are always dealing with non-relativistic DM particles).
Here $Y_{\rm eq}$ denotes an equilibrium comoving density $Y_{\rm eq} = \frac{45}{2 \pi^4}\left(\frac{\pi}{8}\right)^{1/2}\frac{g}{g_{*\rm s}}x^{3/2} e^{-x}$, where $g$ is the number of internal degrees of freedom (equal to 2 both for the fermionic and the scalar DM case). The  actual equilibrium comoving densities for the $^+$ and  $^-$ species are respectively
$ Y^+_{\rm eq} = Y_{\rm eq}\, e^{+\xi}$,   $Y^-_{\rm eq} = Y_{\rm eq}\, e^{-\xi}$,  
 where $\xi = \mu / T$ with $\mu$ being the chemical potential. Since they enter only as the product (see below), the chemical potential disappears from the equations. 
It is also useful to introduce the  parameter $\eta_0=Y^+_0-Y^-_0$, which represents the initial DM -- $\overline{{\rm DM}}$ asymmetry and is related to  $\xi_0$ as $\xi_0 = {\rm arcsinh}(\eta_0/(2 Y_{\rm eq}(x_0)))$.

\subsection{Annihilations only}
\label{sec:annonly}

In the case with annihilations only, the density matrix equation in eq.~(\ref{masterequation}) reduces to
\begin{equation}
\label{eqannonly}
\mathcal{Y}^{\, \prime}(x) = -\frac{s(x)}{x\, H(x)} \left( \frac{1}{2} \Big\{ \mathcal{Y}(x), \Gamma_{\rm a}\, \bar{\mathcal{Y}}(x) \, \Gamma_{\rm a}^\dagger \Big\} - \Gamma_{\rm a} \, \Gamma_{\rm a}^\dagger \, \mathcal{Y}_{\rm eq}^2 \right). 
\end{equation}
The right hand side, in particular with its anti-commutator structure, reproduces the more detailed collision integrals  and once the integral over the phase space of incoming and outgoing particles has been performed, as discussed in~\cite{formalism}. We neglect the effects related to the quantum-statistical distribution of particles (e.g. Fermi-blocking factors). 
Here $\Gamma_{\rm a}$ is a diagonal matrix (actually proportional to the identity in the case at hand) defined in such a way that $\Gamma_{\rm a} \, \Gamma_{\rm a}^\dagger = \langle \sigma v \rangle \, \mathbb{I}$, where $\langle \sigma v \rangle$ is the thermally averaged annihilation cross section. $\langle \sigma v \rangle$ admits the usual expansion in even powers of the velocity $v$ of the DM particles
\begin{equation}
\langle \sigma v \rangle = \sigma_0 + \sigma_1 \langle v\rangle^2 + \mathcal{O}(v^4),
\label{sigmav}
\end{equation}
For simplicity, we will always assume $s$-wave annihilations in the following, which amounts to keep only the first term of the expansion.
$\bar{\mathcal{Y}}$ is the charge-conjugated matrix of $\mathcal{Y}$, i.e. the same quantity as the latter but with the role of particles and antiparticles flipped: $\bar{\mathcal{Y}} =  {\rm CP}^{-1} \cdot \mathcal{Y}\cdot {\rm CP}$, where ${\rm CP} = i \sigma_2 ={\tiny \left( \begin{array}{cc}0 & 1\\ -1 & 0\end{array} \right)}$.
Finally, the matrix $\mathcal{Y}_{\rm eq}^2$ reads $\mathcal{Y}_{\rm eq}^2 =  {\tiny \left( \begin{array}{cc}Y_{\rm eq}^2 & 0\\ 0 & Y_{\rm eq}^2\end{array} \right)}$.

In solving eq.~(\ref{eqannonly}), the off-diagonal components  remain identically zero and the whole information on the evolution of the system is encoded in the equations for the diagonal components $Y^\pm$. Such equations can then be recast in the more familiar Boltzmann form~\cite{ScherrerTurner}:
\begin{equation}
Y^{\pm\, \prime}(x) = -\frac{\langle \sigma v \rangle \, s(x)}{x\, H(x)}\left[Y^+(x)\, Y^-(x)-Y_{\rm eq}^2(x)\right].
\label{Beqannonly}
\end{equation}

\medskip

\begin{figure}
	\parbox[b]{.48\linewidth}{
		\includegraphics[width=\linewidth]{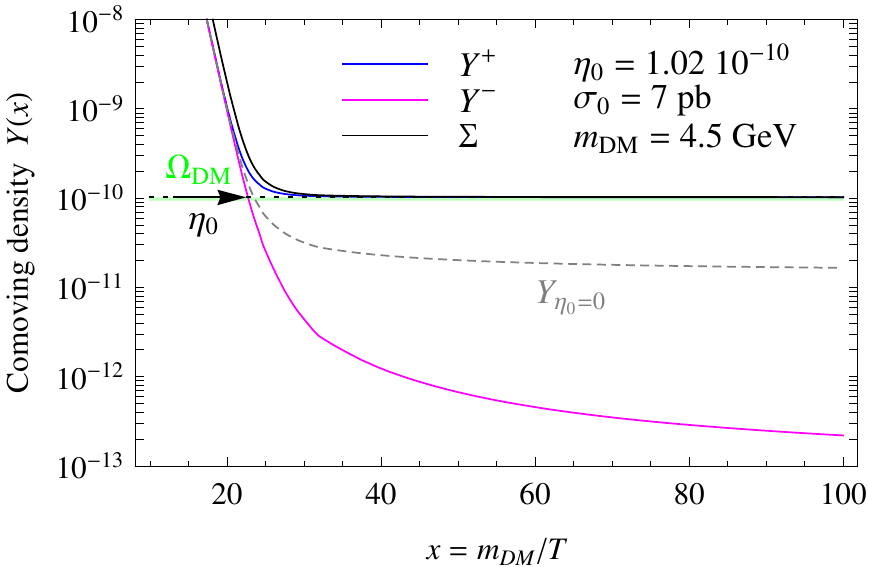}}\hfill
	\parbox[b]{.46\linewidth}{ 
		\includegraphics[width=\linewidth]{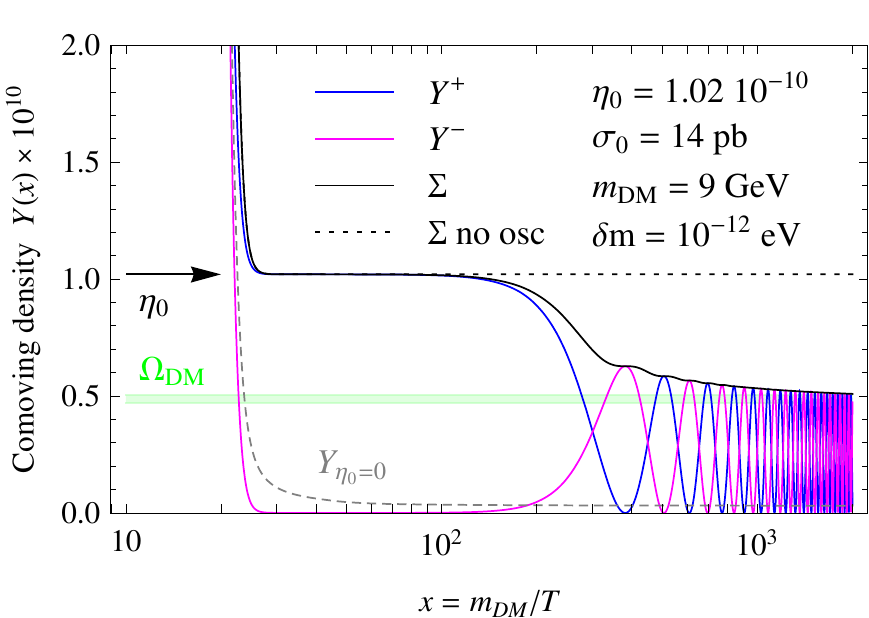}} 
	\parbox[b]{.48\linewidth}{
		\includegraphics[width=\linewidth]{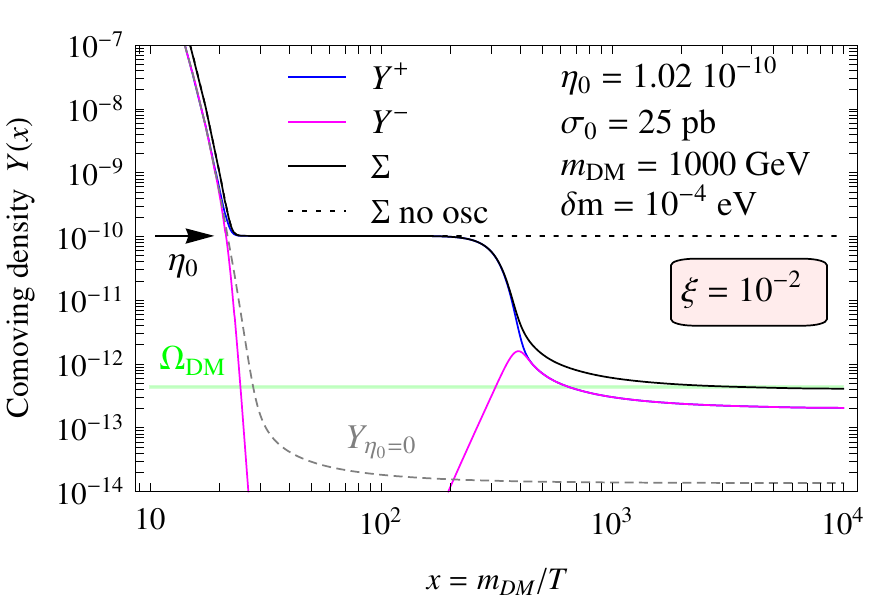}}\hfill
	\parbox[b]{.48\linewidth}{
		\caption{\em \small Illustrative plots of the solutions of the evolution equations in the case of annihilations only (top left panel, discussed in Sec.~\ref{sec:annonly}), annihilations with oscillations (top right panel, Sec.~\ref{sec:ann+osc}) and in the case which includes elastic scatterings (bottom left panel, Sec.~\ref{sec:wscatterings}). The blue (magenta) line represents the comoving population of $n^+$ ($n^-$), the black line their sum. The arrow points to the value of the primordial asymmetry, the green band is the correct relic abundance ($\pm\, 1 \sigma$).}
		\label{figannosc}
	}
\end{figure}

It is now straightforward to solve the equations (\ref{eqannonly}) (or, equivalently, eq.~(\ref{Beqannonly}), as it has been done in~\cite{Graesser:2011wi,Iminniyaz:2011yp}). We show in fig.\ref{figannosc} (upper left panel)
the result in the specific case  $\eta_0 = \eta_{\mbox{\tiny B}} = 1.02 \ 10^{-10}$ (the latter being the value of the baryonic asymmetry, see e.g.~\cite{cosmoDM})~\footnote{Note that we have defined here the quantities $\eta$, for DM and for baryons, in terms of the ratio of the difference of number densities with entropy $s$: $\eta= (n- \bar n)/s$. This notation is not to be confused with the one (sometimes also denoted $\eta$) involving the ratio with the photon number density. In this latter notation, the baryon to photon ratio $(n_{\mbox{\tiny B}} - \bar n_{\mbox{\tiny B}})/n_\gamma \simeq n_{\mbox{\tiny B}}/n_\gamma$ equals the familiar value $6.18 \ 10^{-10}$~\cite{cosmoDM}.} and where we have taken a large annihilation cross section. 
Let us comment on the main qualitative features. At small $x$, the presence of a primordial asymmetry is irrelevant and both comoving densities follow essentially the equilibrium curve. Freeze-out happens when the system runs out of targets, and then the absolute value of $Y^+$ (assumed to be the most abundant species) approaches $\eta_0$: $Y^+$ sits on a plateau while the contribution of $Y^-$ can be neglected. As anticipated, therefore, in this typical aDM configuration the most relevant parameter is the initial asymmetry $\eta_0 = \eta_{\mbox{\tiny B}}$: it sets the asymptotic number density~\footnote{Note that we are assuming that any process changing the DM-number (such as e.g. weak sphalerons, in models in which the DM-number is related to the ordinary baryon number) is already switched off by the time of freeze-out, so that we can consider $\eta_0$ as an actual constant in the subsequent evolution. This could be invalid for very large DM masses ($\gtrsim 10$ TeV), for which freeze-out happens early.} and thus, in order to obtain the correct $\Omega_{\mbox{\tiny DM}}$, forces $m_{\mbox{\tiny DM}}$ to be $\cal{O}$(5 GeV) (4.5 GeV in the plot).

\medskip

For illustration one can also define the sum and the difference of the comoving number densities
\begin{equation}
\Sigma(x) = Y^+(x) + Y^-(x), \qquad \Delta(x) = Y^+(x) - Y^-(x),
\label{YtotDelta}
\end{equation}
In terms of these quantities, the Boltzmann equations read 
\begin{equation}
\left\{ 
\begin{split}
& \Sigma^{\, \prime}(x) = - 2 \frac{\langle \sigma v \rangle \, s(x)}{x\, H(x)}\left[\frac{1}{4} \Bigg(\Sigma^2(x)- \Delta^2(x) \bigg) -Y_{\rm eq}^2(x)\right],\\
& \Delta^{\prime}(x) = 0,
\end{split}
\right.
\label{BeqannonlyYD}
\end{equation}
which clearly shows that  the difference $\Delta$ between the populations remains constant (and equal to the initial condition $\eta_0$); on the other hand, the total population $\Sigma$ of $^+$ and $^-$ particles decreases, due to annihilations. At late times, $Y_{\rm eq}$ is negligible and $\Sigma$ is attracted towards $\Delta =\eta_0$.

\subsection{Oscillations only}
\label{sec:osconly}

We consider next the restricted case in which there are only ${\rm DM} \leftrightarrow \overline{{\rm DM}}$  oscillations in the system, without annihilations nor scatterings with the plasma. Eq.~(\ref{masterequation}) reduces in this case to the simple form
\begin{equation}
\label{eqosconly}
\mathcal{Y}^{\, \prime}(x) = - \frac{i}{x\, H(x)} \Big[\mathcal{H},\mathcal{Y}(x) \Big].
\end{equation}
where $\mathcal{H}$ is the Hamiltonian of the system, which, as discussed in Sec.~\ref{sec:theory}, we parametrize as 
\begin{equation}
\label{Hosconly}
\mathcal{H} =  \left( \begin{array}{cc} m_{\mbox{\tiny DM}} & \delta m \\ \delta m & m_{\mbox{\tiny DM}} \end{array} \right).
\end{equation}
The system of four coupled equations for the individual entries of the matrix $\mathcal{Y}$ can be explicitly solved analytically. The off-diagonal components can be plugged in the equations for the diagonal components $Y^\pm$ and one finds that those correspond to the following familiar Boltzmann equations:
\begin{equation}
\left\{ 
\begin{split}
& Y^{+\, \prime}(x) = -\frac{\Gamma_{\rm osc}(x)}{x\, H(x)} \left[Y^+(x) - Y^-(x)\right],\\
& Y^{-\, \prime}(x) = -Y^{+\, \prime}(x),
\end{split}
\right.
\label{Beqosconly}
\end{equation}
with the same initial conditions as for eq.~(\ref{masterequation}) and where the oscillation rate is defined as  
\begin{equation}
\Gamma_{\rm osc}(x) = \delta m \ \tan\left(\frac{\delta m}{H(x)}\right).
\label{Gammaosc}
\end{equation}
These can also be written in terms of $\Sigma$ and $\Delta$ as
\begin{equation}
\left\{ 
\begin{split}
& \Sigma^{\, \prime}(x) = 0,\\
& \Delta^{\, \prime}(x) = - 2 \frac{\Gamma_{\rm osc}(x)}{x\, H(x)} \Delta(x).
\end{split}
\right.
\label{BeqosconlyYD}
\end{equation}
It is now $\Sigma$ which is constant in time, since oscillations exchange particle with antiparticle but conserve the total number of bodies, while $\Delta(x)$ follows an oscillatory behaviour. 

\bigskip

\begin{figure}[!t]
\begin{center}
\includegraphics[width=0.48 \textwidth]{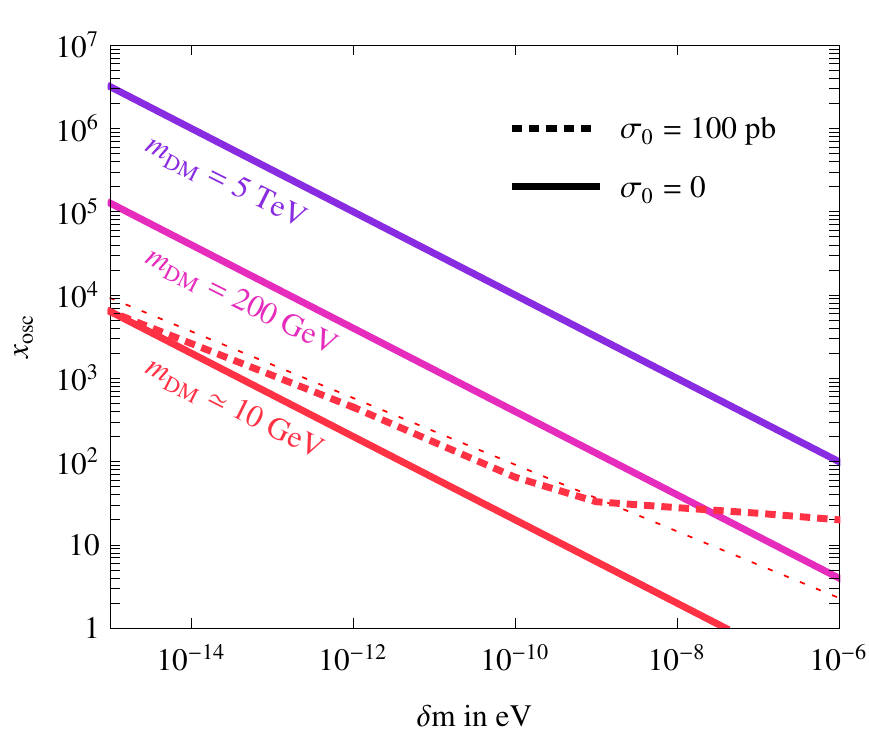} \quad
\includegraphics[width=0.48 \textwidth]{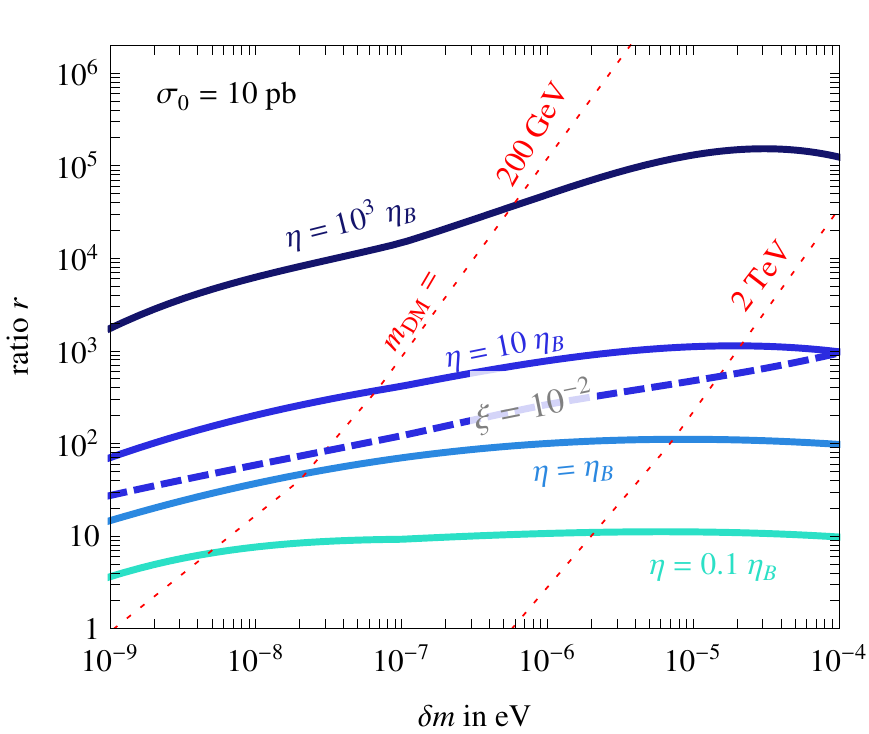}
\caption{\em \small \label{parameterranges} {\bf \em Left panel}: illustration of the approximate relation in eq.~(\ref{xosc}) and eq.~(\ref{xoscann}), i.e. the value of $x$ at which oscillations start as a function of $\delta m$ for a few indicative values of the DM mass. The dotted lines trace the modification to that relation in the case where annihilations are active, see Sec.~\ref{sec:ann+osc}. {\bf \em  Right panel}: graphical illustration of the approximate relation in eq.~(\ref{eq:ratio}), i.e. the efficiency of oscillations in depleting the aDM excess (for definiteness, in the case of no elastic scatterings, i.e. $\xi = 0$, except for the dashed line marked by the label $\xi = 10^{-2}$). The crossings of the diagonal dotted lines with the four solid lines individuate the values of $\delta m$ for which $\Omega_{\mbox{\tiny \rm DM}}$ reproduces the correct abundance, for the indicated values of $m_{\mbox{\tiny  \rm DM}}$.
}
\end{center}
\end{figure}

In the absence of interactions with the plasma, the probability that a DM particle  becomes a 
$\overline{\mbox{DM}}$ particle at time $t$ is simply $P_{\rm osc}^{+-}(t) =\sin^2 \left(\delta m \ t \right)$.
Oscillations start when  $H(x) \lesssim \delta m $ (i.e $T\lesssim \sqrt{\delta m \, M_{\rm Pl}}$). Slightly more precisely, one can define $x_{\rm osc}$ via the condition $\delta m~ x^2 _{\rm osc}/H(m_{\mbox{\tiny DM}}) \simeq 2\pi$, which gives
\begin{equation}
x_{\rm osc} \simeq \left(\frac{8 \pi^3}{90} g_*\right)^{1/4} \frac{1}{\sqrt{M_{\rm Pl}}} \frac{m_{\mbox{\tiny DM}}}{\sqrt{\delta m}} \approx 2 \cdot 10^{-4} \left(\frac{m_{\mbox{\tiny DM}}}{10 \ {\rm GeV}}\right) \left(\frac{{\rm eV}}{\delta m}\right)^{1/2}.
\label{xosc}
\end{equation}
This equation is plotted in Fig.~\ref{parameterranges}, showing  that a large range of possibilities is open, depending on the values of the DM mass and of the $\delta m$ parameter. We will later see how this relation is modified by the presence of annihilations and elastic scatterings.

%
%

\subsection{Combining annihilations and oscillations}
\label{sec:ann+osc}

When combining annihilations and oscillations, the features that we separately highlighted above  overlap: initially the total number of particles decreases due to annihilations; later, when oscillations start, they repopulate $Y^-$ at the expense of $Y^+$ so that annihilations can recouple, thus reducing the sum; as a consequence, in the next `cycle', the total number of $Y^+$ and $Y^-$ subject to oscillations is reduced, i.e. the amplitude of the oscillation also decreases. The amount by which the amplitude of the oscillations decreases is determined by the amount by which the sum of particles at disposal decreases.

\bigskip 

All this is accounted for by eq.~(\ref{masterequation}), that we reproduce here for convenience:
\begin{equation}
\label{eqannosc}
\mathcal{Y}^{\, \prime}(x) =  -i \frac{1}{x\, H(x)} \Big[\mathcal{H},\mathcal{Y}(x) \Big] -\frac{s(x)}{x\, H(x)} \left( \frac{1}{2} \Big\{ \mathcal{Y}(x), \Gamma_{\rm a}\, \bar{\mathcal{Y}}(x) \, \Gamma_{\rm a}^\dagger \Big\} - \Gamma_{\rm a} \, \Gamma_{\rm a}^\dagger \, \mathcal{Y}_{\rm eq}^2 \right). 
\end{equation}
This equation can be recast into a set of coupled Boltzmann-like relations, namely:
\begin{equation}
\left\{ 
\begin{split}
& \Sigma^{\, \prime}(x) = - 2 \frac{\langle \sigma v \rangle \, s(x)}{x\, H(x)}\left[\frac{1}{4} \Bigg(\Sigma^2(x)- \Delta^2(x) - \Xi^2(x) \bigg) -Y_{\rm eq}^2(x)\right],\\
& \Delta^{\prime}(x) = \frac{2i\,\delta m}{x\, H(x)}\, \Xi(x),\\
&\Xi^{\prime}(x) = \frac{2i\,\delta m}{x\, H(x)}\,\Delta(x)~-\frac{\langle \sigma v \rangle \, s(x)}{x\, H(x)}\Xi(x) \Sigma(x).
\end{split}
\right.
\label{Beqannosc}
\end{equation}
where $\Xi$ corresponds to the difference between off-diagonal elements of the density matrix, $\Xi(x)=Y^{+-}(x)-Y^{-+}(x)$. From this, it is clear
that the system cannot be reduced to simple equations for the two functions $\Sigma$ and $\Delta$ (already defined above). In other words, the interplay of coherent and incoherent processes cannot be thoroughly followed by focussing only on the populations of $Y^+$ and $Y^-$, or their sum and difference: one more functional `degree of freedom' is needed.

\medskip

Some insight can anyhow be learnt by considering the (oversimplified) case featuring oscillations and a {\em constant} effective rate of annihilations, denoted $\gamma_{\rm a}$, and neglecting variation with $x$ of the total population $\Sigma$. In this case, by combining the second and third equations in (\ref{Beqannosc}), one arrives at an effective equation $\Delta^\prime \simeq -2\, \delta m^2/\gamma_{\rm a}\, \Delta$, valid in the regime $\gamma_{\rm a} \gg \delta m$. Contrasted with the second of eq.~(\ref{BeqosconlyYD}), this shows that, in presence of annihilations, the difference between the populations dims with a rate proportional to $\delta m^2/\gamma_{\rm a}$, a point to which we will come back later. \label{discussion on deltam2/gamma} However, we stress that this simplification does not allow to include all the features of the system. We will stick to the full equation (\ref{eqannosc}) for the numerical solutions in the following.

\bigskip 

In figure \ref{figannosc} (top right panel) we show the numerical result of eq.~(\ref{eqannosc}) (or, equivalently, eq.~(\ref{Beqannosc})) for a specific illustrative case. Like for the top left example in the same figure, we have again taken $\eta_0 = \eta_{\rm B}$, but here the population $Y^+$ sits only temporarily on the plateau determined by $\eta_0$. With a value of $\delta m = 10^{-12}$ eV, oscillations start at $x \sim 300$ and we see $Y^-$ being repopulated. Given the relatively large annihilation cross section $\sigma_0 = 14$ pb, annihilations can then promptly resume and the total population $\Sigma$ decreases. In the later stages, $\Sigma$ goes through a rapid series of plateaux and drops, until it rests on its asymptotic value, determined by the freeze-out of annihilations. One can therefore have a final $Y_\infty \ll \eta_0$ and obtain the same $\Omega_{\mbox{\tiny DM}} = \rho_{\mbox{\tiny DM}}/\rho_{\rm crit} = m_{\mbox{\tiny DM}} Y_\infty s /\rho_{\rm crit}$ with a large DM mass with respect to the standard aDM case. In other words, this example illustrates how, as anticipated in the introduction,  $\Omega_{\mbox{\tiny DM}}$ is no longer determined by $\eta_0$ but by the combination of different parameters $\eta_0, m_{\mbox{\tiny DM}}, \delta m, \sigma_0$. 
We will discuss several illustrative choices for these in Section~\ref{sec:results}. 
Note that our formalism  allows to follow in detail the oscillatory pattern (evident at large $x$ in fig.~\ref{figannosc}b). In other approaches in the literature only an effective average of oscillations has been employed (see e.g.~\cite{Falkowski:2011xh,Cui:2011qe}). While this may be enough for an estimate of the effect or for the late $x$ behaviour, it may miss the details at the starting-up of oscillations.

\bigskip 

Another non trivial effect of the interplay between annihilations and oscillations has to do with the moment of the start of oscillations. While in a purely coherent (albeit expanding) system with only oscillations, as the one we considered in Sec.~\ref{sec:osconly}, the conversions start at a $x_{\rm osc}$ determined via eq.~(\ref{xosc}), the addition of annihilations breaks such coherence and effectively delays the picking up of oscillations. 
In top right panel fig.~\ref{figannosc} the effect is barely visible (namely,  $x_{\rm osc}$ equals $\sim 300$ or so, instead of  $x_{\rm osc} \sim 200$ as it would be dictated by eq.~(\ref{xosc})), but for larger values of the $\langle \sigma v \rangle$ parameter the suppression and delay of oscillations becomes more important. 
In terms of the effective simplification discussed below eq.~(\ref{Beqannosc}), where the relevant time scale is now $\delta m^2/\gamma_{\rm a}$, we obtain that oscillations start when 
\begin{eqnarray}
x_{\rm osc, ann} & \simeq & \left( \frac{H_m\, \gamma_{\rm a}}{2\, \delta m^2} \right)^{1/2} \simeq \left( \frac{H_m \, \sigma_0 \, s_m \, \eta_0/2}{\delta m^2} \right)^{1/5} \nonumber \\
 &\approx & 12\,\left(\frac{m_{\mbox{\tiny DM}}}{100\,{\rm GeV}} \right)\,\left( \frac{10^{-7}\, {\rm eV}}{\delta m} \right)^{2/5}\, \left( \frac{g_{*{\rm s}}}{10}\,\sqrt{\frac{g_{*}}{10}}\,\frac{\sigma _0}{1\,{\rm pb}}\,\frac{\eta _0}{\eta_{\mbox{\tiny B}}}\right)^{1/5},
\label{xoscann}
\end{eqnarray}
where $s_m = s(x=1)$, in analogy with $H_m$.
In fig.~\ref{parameterranges} (left panel) we also report, for the specific case of $m_{\mbox{\tiny DM}} = 10$ GeV, the effective value of $x_{\rm osc,ann}$ for an annihilation cross section of $\sigma_0 = 100$ pb. We plot the value as predicted by eq.~(\ref{xoscann}) (thin dotted line) and as determined numerically (thick dotted line).

\subsection{Including elastic scatterings}
\label{sec:wscatterings}

Dark Matter (and antiDM) particles travel through the dense primordial plasma and elastically scatter on it via 
${\rm DM}\, {\rm SM} \rightarrow {\rm DM}\, {\rm SM}$ processes, where `SM' denotes any Standard Model particle that is abundant enough in the plasma, i.e. essentially relativistic species. 
This affects the evolution of the system in two main ways (we follow closely for this discussion the case of neutrino propagation in matter, see e.g.~\cite{neutrinoreviews}): (i) an effective matter potential $V$ is generated by the coherent interactions and enters in the commutator part of the density matrix equation;
(ii) the incoherent scatterings give rise to a rate of interactions $\gamma_{\rm s}$ entering in the anti-commutator part.

The whole system is therefore now described by eq.~(\ref{masterequation}) with all pieces included and where
\begin{equation}
\label{HandGammas}
\mathcal{H} =  \left( \begin{array}{cc} m_{\mbox{\tiny DM}} +V(x) + \Delta V(x) & \delta m \\ \delta m & m_{\mbox{\tiny DM}}  +V(x) \end{array} \right) \qquad {\rm and} \qquad \Gamma_{\rm s} =  \left( \begin{array}{cc} \gamma_{\rm s} & 0 \\ 0 & \gamma_{\rm s} \end{array} \right).
\end{equation}
The common terms on the diagonal of $\mathcal{H}$ of course do not have any effect on oscillations, while the difference $\Delta V$ does. 
$\Delta V$ represents the effective energy shift of DM versus  $\overline{\rm DM}$ induced by the baryon asymmetry of the medium. Effectively, it leads  to a non-maximal mixing angle, thus reducing the oscillation probability 
in the vacuum $P_{\rm osc}^{+-}$ by a factor $4 \delta m^2/(4 \delta m^2+\Delta V^2)$.
 For simplicity we assume that $\delta m$ is not affected by the medium.

The explicit form of $\Delta V$ and $\gamma_{\rm s}$ depends on the specific interactions of DM with the plasma. Since we are mainly interested in the case of Weakly Interacting dark matter, we mimic them from those of neutrinos. 
An important point to notice, however, is that the same scatterings we are considering here are also those that would produce signals in DM direct detection experiments, i.e. nuclear or electron recoils in low background set-ups. In order to be consistent with direct detection experiments, therefore, we assume that the DM coupling with matter is suppressed with respect to the weak coupling. On the basis of these observations, we take
\begin{equation}
\label{DeltaVandgammas}
\Delta V =  \xi \ \sqrt{2} \, G_{\rm F} \, \eta_{\mbox{\tiny B}} \, \big(g_{*\rm s}(x) -2\big) \, n_{\rm bos}  \qquad {\rm and} \qquad  \gamma_{\rm s} =  \xi^2 \ \frac{45}{\pi^3} \zeta (5)\, G_{\rm F}^2\, \big(g_{*\rm s}(x) -2\big) \frac{m_{\mbox{\tiny DM}}^5}{x^5},
\end{equation}
where $G_{\rm F}$ is the Fermi constant, $n_{\rm bos} = 1/\pi^2 \,\zeta(3) \, m_{\mbox{\tiny DM}}^3/x^3$ is the number density per degree of freedom of relativistic bosons and $\zeta(n)$ is the Riemann zeta function of $n$. 
In the equations above the presence of the factor $(g_{*\rm s}(x) -2)$ is due to the fact that we take into account that WIMP DM scatters on all the relativistic degrees of freedom (counted by $g_{*\rm s}(x)$) except for photons. Also, by using $\eta_{\mbox{\tiny B}}$ in the expression for $\Delta V$, we are implicitly assuming that all relativistic SM species share the same asymmetry, equal to the baryonic one.\footnote{Notice that no term proportional to the DM asymmetry itself is present, since the DM and $\overline{\rm DM}$ population is Boltzmann suppressed in the regimes of our interest. As a consequence, there is no feedback of the evolution of the DM asymmetry into $\Delta V$  (such a feedback is instead present in the case of relativistic neutrinos in the Early Universe).} 

The parameter $\xi$ expresses the suppression of the Fermi constant due to the fainter DM coupling with matter, as discussed above. Direct detection experiments impose 
$\xi \lesssim 10^{-2}$.
On the other hand, one can check that for $\xi \ll 10^{-3}$ the presence of scatterings has essentially no effect on the system. We will therefore consider in this work two main cases: 
\begin{itemize}
\item[(a)] $\xi \equiv 0$ (i.e. no scatterings), in which case the system reduces to the one discussed in Sec.~\ref{sec:ann+osc}; this scenario makes more evident the effect of oscillations and maximizes their importance.
\item[(b)] $\xi = 10^{-2}$, the maximum allowed value, which makes elastic scatterings, besides annihilations and oscillations, important for the evolution of the DM and $\overline{\rm DM}$ populations. For  large scattering, oscillations are damped, as in the case of standard neutrino mixing in the early universe. 
\end{itemize}
We stress again that eq.s~(\ref{DeltaVandgammas}) are just choices made for definiteness, since we lack a detailed model of the interactions of DM with SM matter. 
For instance, if the DM particle couples only to other dark states which ultimately decay to SM ones, $\Delta V$ and $\gamma_{\rm s}$ are expected to be small. For another instance, if DM is leptophilic and couples only to leptons, then the relevant asymmetry $\eta$ in $\Delta V$ would be the leptonic one, which is poorly constrained. Our formalism allows us to explore most of the possible parameter space while remaining model-independent.

Finally, note that in order to reproduce the  correct  physical system with the last anti-commutator in eq.~(\ref{masterequation}) (which is an approximation to more detailed expressions of the collision integrals~\cite{formalism}), one needs to forbid the terms proportional to $\gamma_{\rm s}$ in the equations for the diagonal components of $\mathcal{Y}$, as commonly done in the literature. 
This guarantees that elastic scatterings do not have the effect of depleting the populations of $Y^+$ and $Y^-$.

\medskip

As done in the previous Subsections, one can derive a set of Boltzmann-like equations from the matrix equation in eq.~(\ref{masterequation}) with eq.~(\ref{HandGammas}). They read 
\begin{equation}
\left\{ 
\begin{split}
& \Sigma^{\, \prime}(x) = - 2 \frac{\langle \sigma v \rangle \, s(x)}{x\, H(x)}\left[\frac{1}{4} \Bigg(\Sigma^2(x)- \Delta^2(x) - \Xi^2(x) -\Pi^2(x) \bigg) -Y_{\rm eq}^2(x)\right],\\
& \Delta^{\prime}(x) = \frac{2i\,\delta m}{x\, H(x)}\, \Xi(x),\\
&\Xi^{\prime}(x) = \frac{2i\,\delta m }{x\, H(x)} \,\Delta(x)~-\frac{i\, \Delta V}{x\, H(x)}\, \Pi(x)~-\frac{\gamma_{\rm s}}{x\, H(x)}~\Xi(x)-\frac{\langle \sigma v \rangle \, s(x)}{x\, H(x)}\Xi(x) \Sigma(x), \\
&\Pi^{\prime}(x) = -\frac{i \, \Delta V}{x\, H(x)}\, \Xi(x) -\frac{\gamma_{\rm s}}{x\, H(x)}~\Pi(x) .
\end{split}
\right.
\label{Beqannoscscatt}
\end{equation}
Yet one more functional degree of freedom coupled to the others, the function $\Pi(x) = Y^{+-}(x)+ Y^{-+}(x)$, has to be introduced. The interplay of the coherent and incoherent processes (annihilations and scatterings) can thus be thoroughly followed by using the full density matrix formalism, either recast in the form of eq.~(\ref{Beqannoscscatt}) or, more conveniently, in the form of eq.~(\ref{masterequation}), to which we will adhere in the following.

\medskip

In order to understand qualitatively the impact of adding incoherent scatterings on the evolution of the populations of DM particles and antiparticles, we can consider the (oversimplified) case of a system featuring oscillations and a {\em constant} $\gamma_{\rm s}$. We neglect $\Delta V$ and we switch off annihilations for simplicity. In this case the matrix equation in eq.~(\ref{masterequation}) schematically reads $\mathcal{Y}^{\, \prime} =  -i/(x\, H) \Big([\mathcal{H},\mathcal{Y} ] - \left\{  \Gamma_{\rm s} 
, \mathcal{Y}\right\} \Big)$. Proceeding in the same way as discussed in Sec.~\ref{sec:osconly}, this equation can be recast into the same pair of coupled Boltzmann equations in eq.~(\ref{Beqosconly}), {\em but} with a more complicated $\Gamma_{\rm osc} = 2\, \delta m^2 / (\gamma_{\rm s}+\omega\, {\rm coth}(\omega/2H(x))$, where $ \omega = \sqrt{\gamma_{\rm s}^2-4\, \delta m^2}$. It is then straightforward to recognize two limits. If elastic scatterings are negligible ($\gamma_{\rm s} \ll \delta m$) then $\Gamma_{\rm osc} \to \delta m \, \tan(\delta m/H(x))$, reducing the system to the case with pure oscillations discussed in Sec.~\ref{sec:osconly}. If instead elastic scatterings are dominant ($\gamma_{\rm s} \gg \delta m$), then at late times $\Gamma_{\rm osc}$ approaches a constant value $\Gamma_{\rm osc} \to 2\, \delta m^2 /\gamma_{\rm s}$. In this situation, the eq.s~(\ref{Beqosconly}) describe a system of $Y^+$ and $Y^-$ densities that are driven, with a strength determined by $\Gamma_{\rm osc} =  2\, \delta m^2 /\gamma_{\rm s}$, one towards the other. In other words, oscillations are damped away and the comoving densities tend asymptotically to their average value. 
Note that in this case, we recover an equation which is often used in the literature
e.g. in \cite{Falkowski:2011xh,Cui:2011qe,Abazajian:2001nj,Dodelson:1993je}:  $Y^{+\, \prime}(x) = -\frac{\langle P \rangle\, \gamma_{\rm s}}{x\, H(x)} \left[Y^+(x) - Y^-(x)\right]$, namely the transfer rate is just the averaged oscillation probability, $\langle P \rangle=\gamma_{\rm s} \int_0^{\infty} dt \, e^{-\gamma_{\rm s}\,t}\sin^2{\delta m \ t}$, multiplied by the interaction rate.
 In the full case, which includes $x$-dependent scattering rates and also annihilations, a comparably simple analytic understanding is not possible, but these general features are preserved, as we move to illustrate next. 

\bigskip

In figure \ref{figannosc} (bottom left panel) we show the outcome of the numerical resolution of eq.~(\ref{masterequation}) with eq.~(\ref{HandGammas}), for a specific illustrative case. As in the previous examples, we have again taken $\eta_0 = \eta_{\rm B}$. In this case, despite the  large $\delta m$, oscillations do not pick up until late, as they are suppressed by the incoherent scatterings. Hence the total density of DM sits for a long time on the familiar plateau. When oscillations do rise (and annihilations restart), barely one oscillating cycle can be seen before they are damped and $Y^+$ and $Y^-$ settle on their asymptotic common value. This illustrates how the inclusion of incoherent scatterings opens a quite different regime: oscillations still have the role of re-symmetrizing the populations allowing for a partial wash-out of the frozen asymmetry, but the ranges of parameters involved are different.

\medskip

We conclude this section by commenting on the quantitative delay of the start of oscillations due to the presence of elastic scatterings. Following the same arguments as in Sec.~\ref{sec:ann+osc}, it is easy to see that now one has approximately 
\begin{eqnarray}
x_{\rm osc, scatt} & \simeq & \left( \frac{H_m\, \gamma_{\rm s}}{2\, \delta m^2} \right)^{1/2} \simeq \left( \frac{H_m \, \gamma_{\rm s}(x=1)}{\delta m^2} \right)^{1/7} \nonumber \\ 
& \approx & 130\,\left(\frac{m_{\mbox{\tiny DM}}}{100\,{\rm GeV}} \right)\,\left( \frac{10^{-7}\, {\rm eV}}{\delta m} \right)^{2/7}\, \left(\left(\frac{g_{*}}{10}\right)^{3/2}\,\frac{\xi}{10^{-2}}\right)^{1/7}.
\label{xoscannscatt}
\end{eqnarray}

\section{Results}
\label{sec:results}

\begin{figure}[!t]
\begin{center}
\includegraphics[width=0.49 \textwidth]{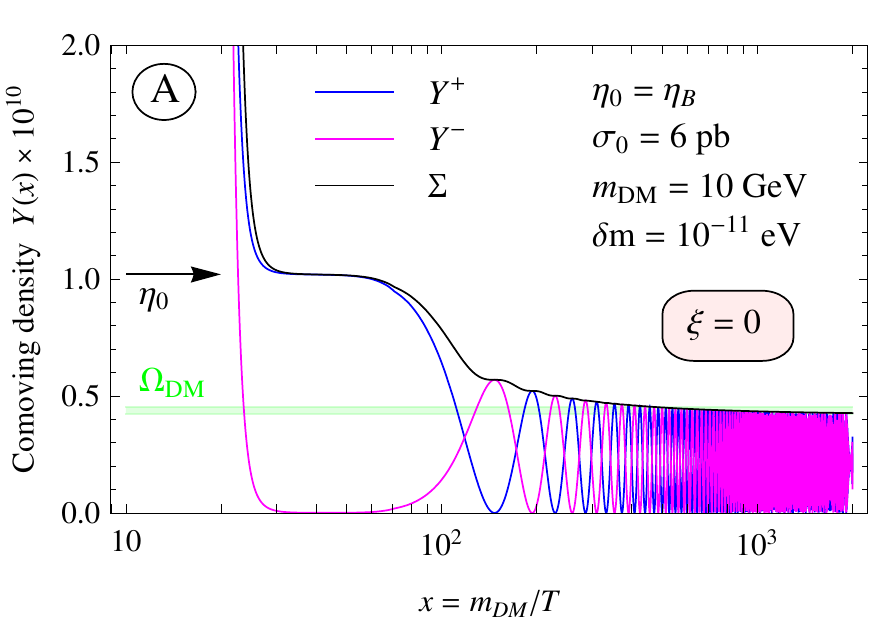} \hfill
\includegraphics[width=0.49 \textwidth]{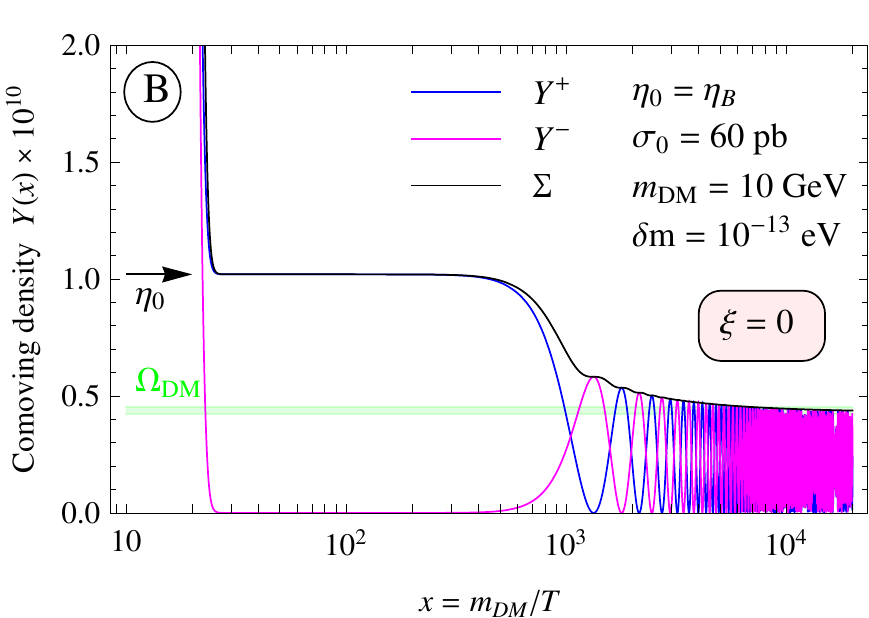}\\
\includegraphics[width=0.49 \textwidth]{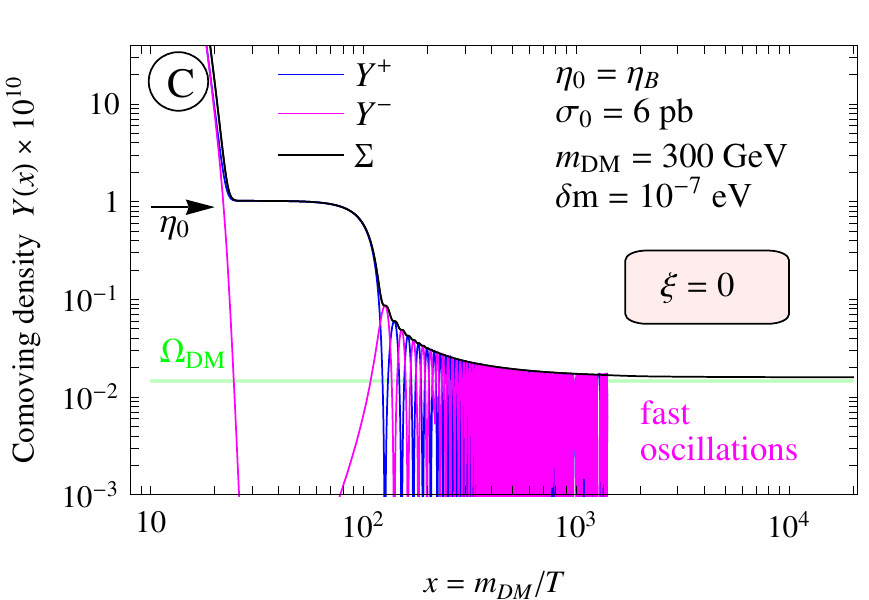} \hfill
\includegraphics[width=0.49 \textwidth]{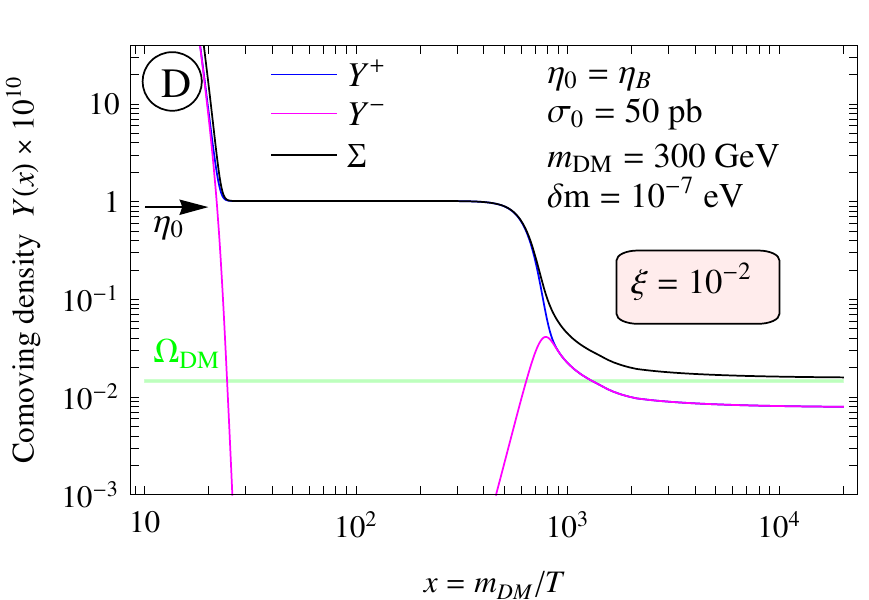}\\
\includegraphics[width=0.49 \textwidth]{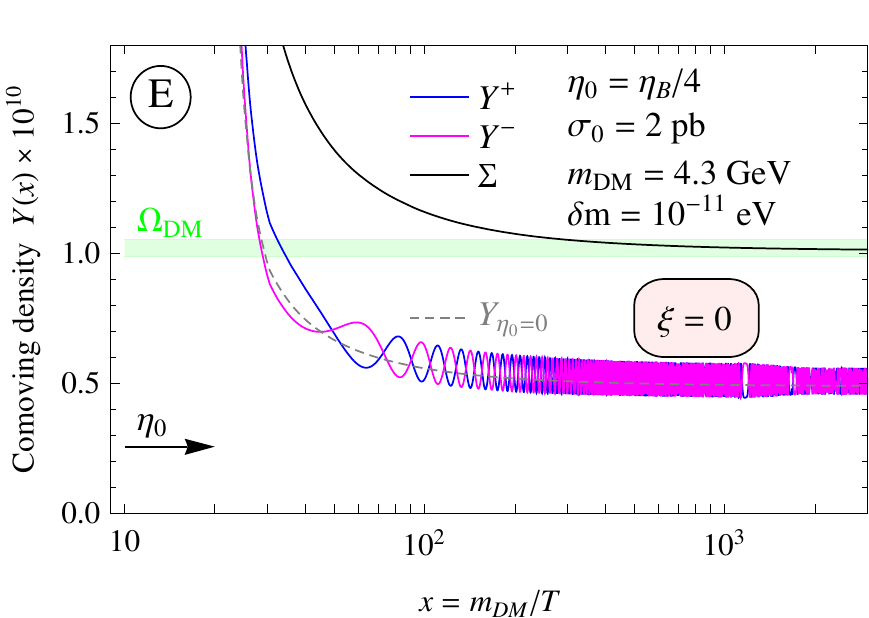} \hfill
\includegraphics[width=0.49 \textwidth]{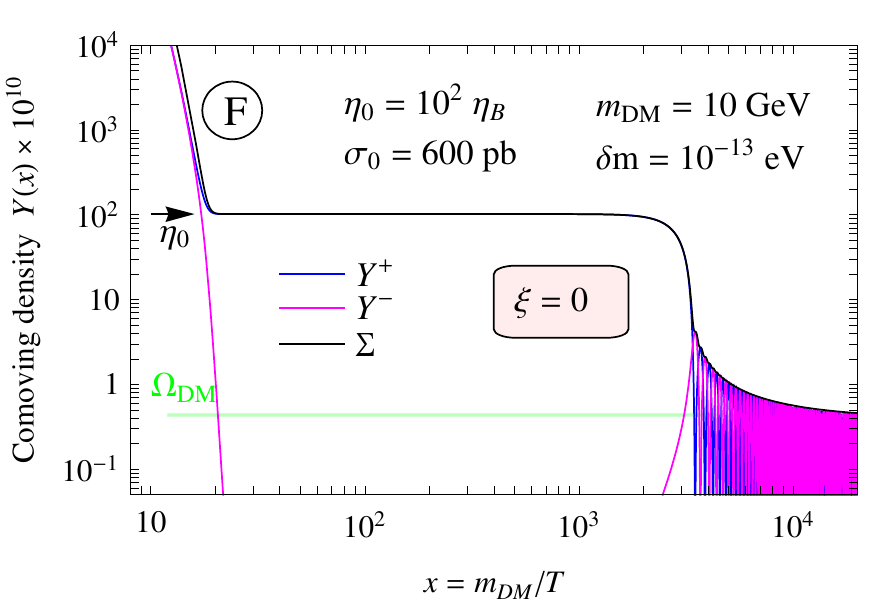}
\caption{\em \small \label{results} 
Some illustrative cases of the time evolution of the populations of DM particles and antiparticles. Notations are like in fig.~\ref{figannosc}, i.e. the blue (magenta) line represents the comoving population of $n^+$ ($n^-$), the black line their sum. The arrow points to the value of the primordial asymmetry, the green band is the correct relic abundance ($\pm\, 1 \sigma$). Notice that some plots have linear scale while other have logarithmic ones, depending on structure which is necessary to show. See text for more details.}
\end{center}
\end{figure}

We now illustrate with some more examples the physics involved in the solutions of the density matrix equations discussed above by varying the parameters $m_{\mbox{\tiny DM}}$, $\sigma_0$, $\eta_0$, $\delta m$ and also $\xi$. 
In fig.~\ref{results} we show the evolution of the comoving dark matter number density in the following cases:
\begin{itemize}
\item[$\circ$] Case A corresponds to choices similar to those discussed in Sec.~\ref{sec:ann+osc} and already adopted for fig.~\ref{figannosc}a and is reported here for the sake of comparing with the following cases.
\item[$\circ$] In Case B we keep the same $m_{\mbox{\tiny DM}}$ as in A, but we adopt a much smaller $\delta m$. The comoving population of DM therefore sits for a longer time on the plateau determined by the initial asymmetry $\eta_0$. However, when oscillations eventually start, annihilations (which have a larger cross section of 60 pb) recouple and can lead to the correct relic abundance. Case B displays therefore the same physics as in A, but delayed in time. Pursuing along this direction, long plateaux can be obtained: an even smaller $\delta m$ would push the start of oscillations further away and a larger $\sigma_0$ would be needed to keep annihilations active that late. 
\item[$\circ$] In case C, we keep instead the same annihilation cross section as in A, but we move to a higher, roughly weak-scale value of the DM mass,  $m_{\mbox{\tiny DM}} =$ 300 GeV. The correct relic abundance is achieved by starting oscillations  earlier than in A, i.e. by choosing a much larger $\delta m$.   
\item[$\circ$] Case D corresponds to same situation as C (in terms of $m_{\mbox{\tiny DM}}$ and $\delta m$), except that now we include elastic scatterings ($\xi = 10^{-2}$). The effect of incoherent scatterings that delay and damp the oscillations is very much apparent with respect to case C. A larger cross section is needed to keep the annihilations active at late times and thus reach the right abundance.  
\item[$\circ$] Case E, on the other hand, illustrates a situation which is similar to the standard thermal freeze-out case, despite the presence of an initial asymmetry and of oscillations. $Y^+$ and $Y^-$ oscillate around the standard solution (in dashed gray, computed for one species only). We have assumed, for this case, a smaller initial asymmetry.
\item[$\circ$] Case F corresponds to a situation in which a very large initial asymmetry (equal to $10^2 \, \eta_B$) is assumed. Having adopted a small $\delta m$, oscillations start late but nevertheless they eventually bring the abundance to the right value. Like for case B, therefore, the comoving density spends a long time on a value which is, in this case, much larger than the final one.
\end{itemize}

\begin{figure}[!tp]
\begin{center}
\includegraphics[width=0.49 \textwidth]{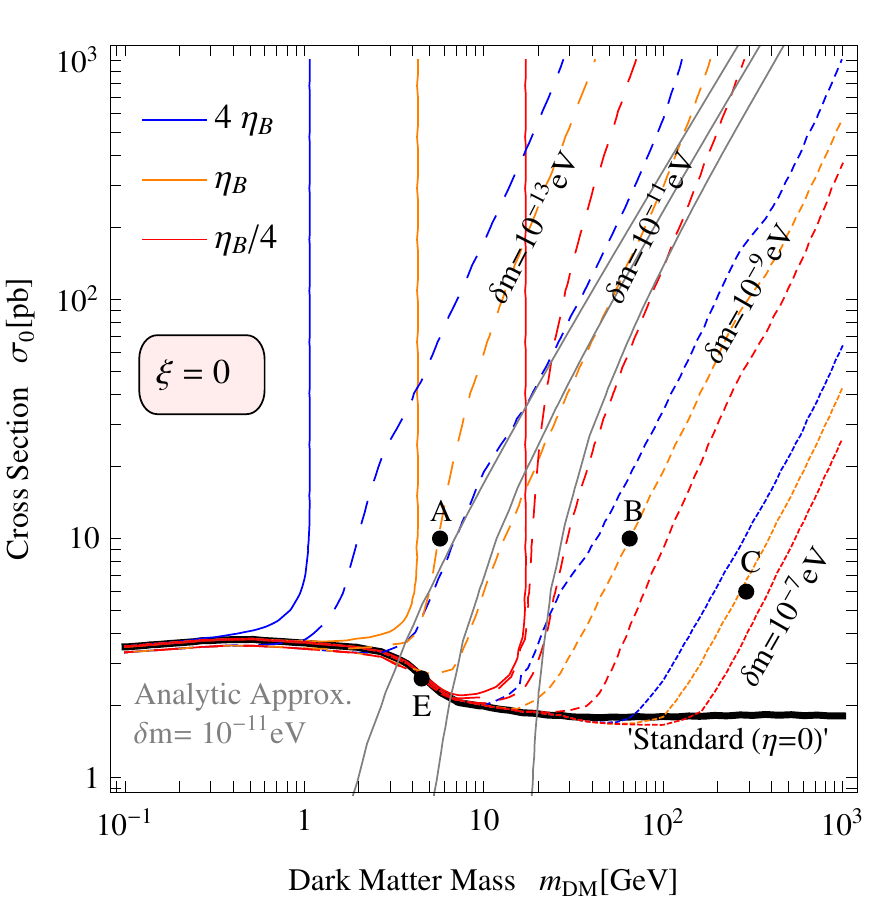}
\includegraphics[width=0.49 \textwidth]{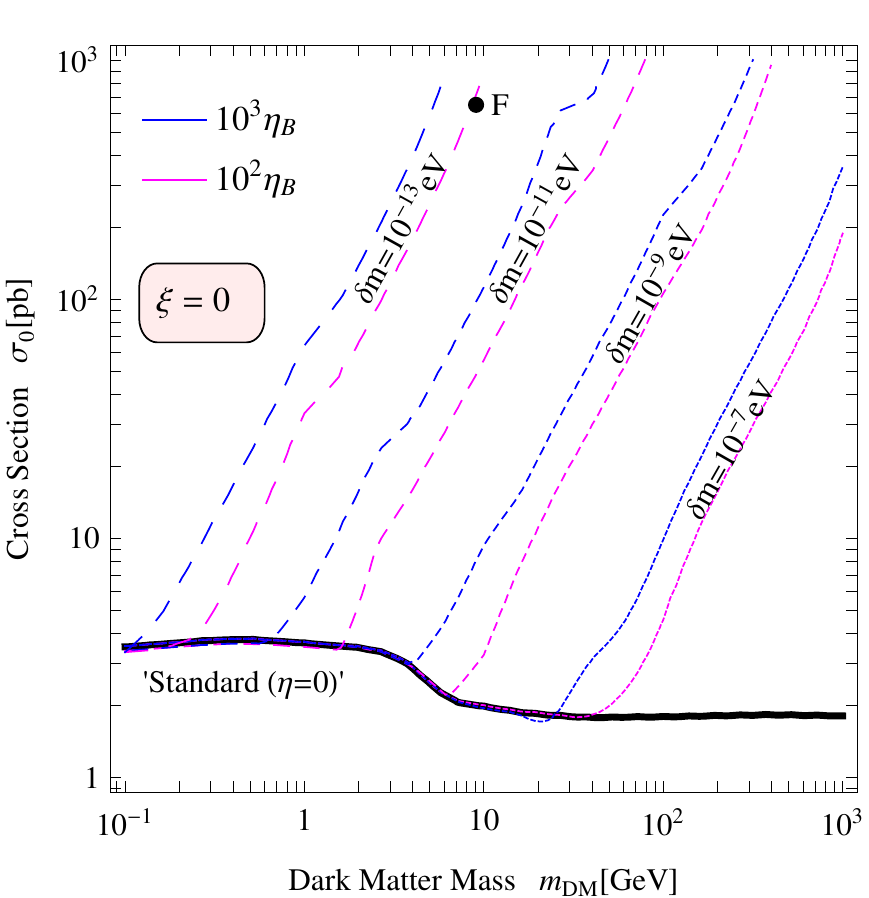} \\
\includegraphics[width=0.49 \textwidth]{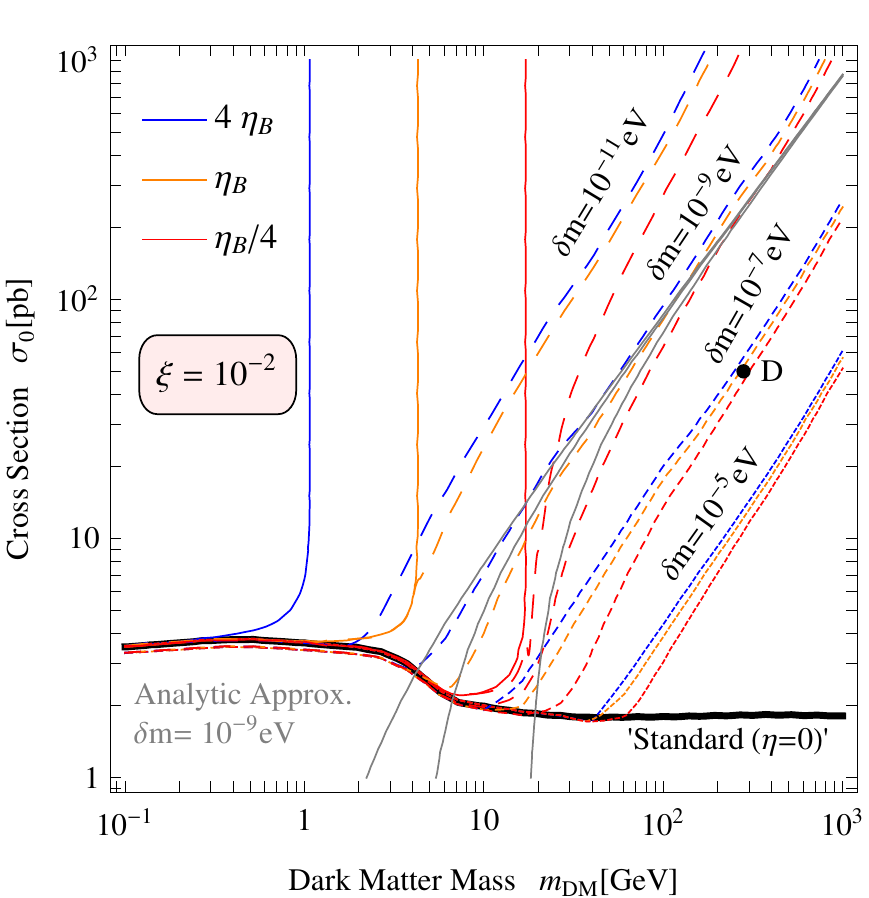}
\includegraphics[width=0.49 \textwidth]{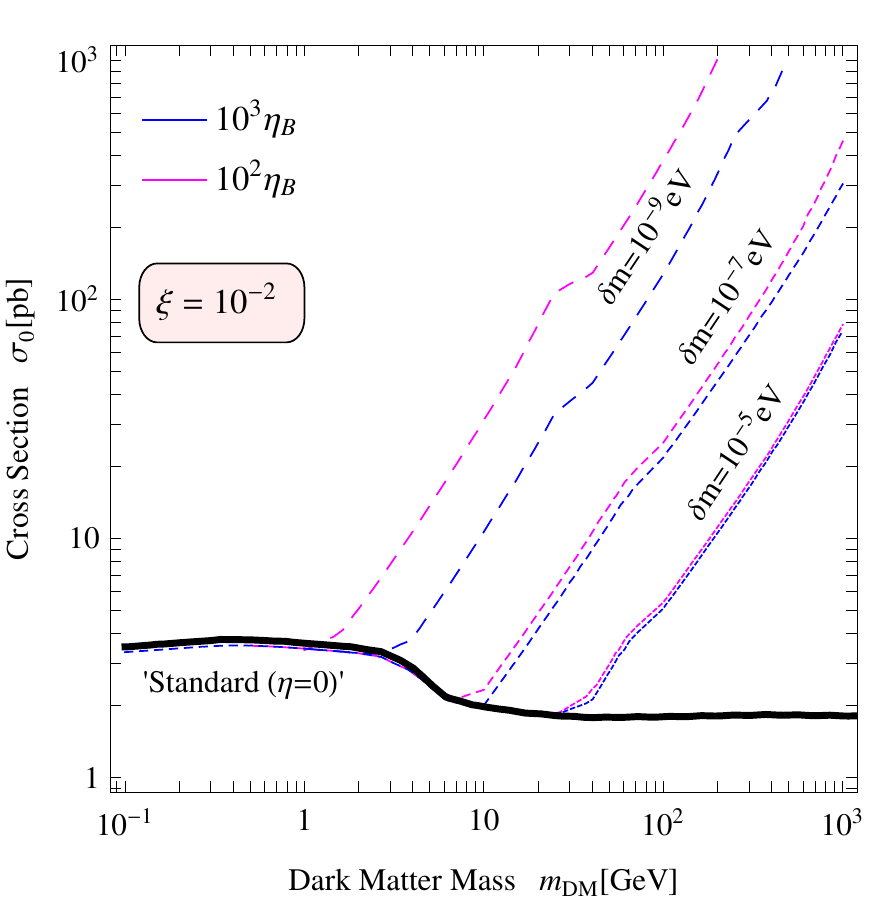} \\
\caption{\em \small \label{paramspace} Contour lines along which a correct $\Omega_{\mbox{\rm \tiny DM}}h^2$ can be obtained, for various values of the initial asymmetry $\eta_0$ (various colors) and several values of the oscillation parameter $\delta m$ (labelled lines marked by different dashings).
 The solid thick black line at the bottom represents the standard case ($\eta = 0, \delta m = 0$). The labelled points (A to F) refer to the cases shown in Fig.~\ref{results}. 
{\bf \em Top panels}: Oscillations and annihilations only, i.e. with $\xi = 0$. 
{\bf \em Bottom panels}: Adding elastic scatterings, i.e. with $\xi = 10^{-2}$. 
The {\bf \em left panels} consider initial asymmetries equal or close to the baryonic one. The {\bf \em right panels} focus on large initial asymmetries. The faint gray lines correspond to the semi-analytic approximations in eq.~(\ref{anapproximationA}) and eq.~(\ref{anapproximationB}).}
\end{center}
\end{figure}

We systematize and summarize our results by showing the contour lines corresponding to the correct DM abundance in the $(m_{\mbox{\tiny DM}},\sigma_0)$ plane in Fig.~\ref{paramspace}. The orange solid line (labelled $\eta_{\mbox{\tiny B}}$) corresponds to the standard aDM scenario. By changing the values of $\delta m$ and $\eta_0$ we can open much more of the parameter space, towards larger $m_{\mbox{\tiny DM}}$ and larger $\sigma_0$. 

\bigskip

Before we move to discuss the constraints on this same parameter space, we want to provide  some approximate analytic expressions that help in identifying the general features of the solutions. 

First, we want to obtain an estimate of the asymptotical value of $\Sigma = Y^++Y^-$, which determines the DM abundance today via $\Omega_{\mbox{\tiny DM}} (x \to \infty) = m_{\mbox{\tiny DM}}\  \Sigma(x \to \infty) \ s /\rho_{\rm crit}$, where $s$ here denotes the entropy density today.
We focus first on the case with elastic scatterings, $\xi = 10^{-2}$, but neglecting $\Delta V$. By solving in an approximate semi-analytic way the system of equations, we are able to obtain an expression for $\Omega_{\mbox{\tiny DM}}  (x \to \infty)$ as a function of all the parameters of the system. It reads
\begin{equation}
\Omega_{\mbox{\tiny DM}} (x \to \infty) \simeq \frac{m_{\mbox{\tiny DM}} \, s}{\rho_{\rm crit}} \eta_0 \left( 1+ \frac{1}{12.2} \frac{g_{* {\rm s},\infty}}{g_{*,\infty} ^{4/7}} \left( \pi^{2}~\frac{\delta m^{2}~M_{\rm Pl}^{8} }{\xi^2 \, \gamma_{{\rm s},0}}\right)^{1/7}\, {\rm Gamma}\left[ \frac{6}{7} \right] \, \sigma_0 \, \eta_0 \right)^{-1},
\label{anapproximationA}
\end{equation}
where $\gamma_{{\rm s},0} = 45/\pi^3\, \zeta(5)\, G_{\rm F}^2\, (g_{*{\rm s}}-2)$ corresponds to the normalization factors of the scattering rate $\gamma_{\rm s}$ in eq.~(\ref{DeltaVandgammas}) and $g_{*({\rm s}),\infty}$ denotes quantities evaluated today.
This approximation holds when, before the onset of oscillations, all $Y^-$ annihilate and the energy density of the Universe is dominated by $Y^+ \sim \eta_0$. It therefore becomes unreliable for large $\delta m$. 
In addition, for large annihilation cross sections (indicatively $\sigma_0 > 100$ pb) the manipulations leading to eq.~(\ref{anapproximationA}) do not hold and thus this approximation fails.
Within this region of validity, we can readily identify two limits. In the limit of no oscillations ($\delta m \rightarrow 0$),  eq.~(\ref{anapproximationA}) reduces to the usual expression in Asymmetric Dark Matter scenarios:
\begin{equation}
\Omega_{\mbox{\tiny DM}} (x \to \infty) \simeq \frac{m_{\mbox{\tiny DM}} \, s}{\rho_{\rm crit}}\, \eta_0. 
\end{equation}
At the other extreme, when the second term in brackets in eq.~(\ref{anapproximationA}) dominates, we have 
\begin{equation}
\Omega_{\mbox{\tiny DM}} (x \to \infty) \simeq \frac{m_{\mbox{\tiny DM}} \, s}{\rho_{\rm crit}} \frac {1}{\frac{1}{12.2}\frac{g_{*{\rm s},\infty}}{g_{*,\infty} ^{4/7}} \left(\pi^{2}~\frac{\delta m^{2}~M_{\rm Pl}^{8} }{  \xi^2 \, \gamma_{{\rm s},0}   }\right)^{1/7}\, {\rm Gamma} \left[ \frac{6}{7} \right] \, \sigma_0} .
\end{equation}
In this limit, the dependence on $\eta_0$ cancels out. Indeed, in fig.~\ref{paramspace}, lower left panel, we observe a degeneracy of the curves corresponding to different initial asymmetries, for the largest $\delta m$ values.
For the case without elastic scatterings ($\xi = 0$), in a similar way we obtain the equivalent of eq.~(\ref{anapproximationA}):
\begin{equation}
\Omega_{\mbox{\tiny DM}} (x \to \infty) \simeq \frac{m_{\mbox{\tiny DM}} \, s}{\rho_{\rm crit}} \eta_0 \left( 1+ \frac{1}{5.1} \frac{g_{*{\rm s},\infty}^{4/5}}{g_{*,\infty}^{3/5}} \left( \frac{\delta m^{2}~M_{\rm Pl}^{6} }{\pi }\right)^{1/5}\, {\rm Gamma} \left[ \frac{4}{5} \right] \, \left(\sigma_0 \, \eta_0\right)^{4/5} \right)^{-1}.
\label{anapproximationB}
\end{equation}
In this case, the asymptotic value of $\Omega_{\mbox{\tiny DM}}$ does carry a more significant residual dependence on $\eta_0$, which indeed we see in figure \ref{paramspace} (upper left panel).
We superimpose the contours determined by eq.~(\ref{anapproximationA}) and eq.~(\ref{anapproximationB}) to the numerical contours in fig.~\ref{paramspace}, for one choice of $\delta m$. We see that the agreement is good within the ranges of validity. 

\medskip

As a second point, we want to quantify the amount by which the parameter space of Asymmetric Dark Matter opens up due to the introduction of oscillations. For this purpose we can define the ratio
\begin{equation}
\label{eq:ratio}
r_{\delta m} (\sigma_0,\eta_0) \equiv \frac{\left. \Omega_{\mbox{\tiny DM}} h^2 \right|_ {\delta m=0}}{\Omega_{\mbox{\tiny DM}} h^2}
\end{equation}
where the numerator expresses the DM abundance that would occur in a standard aDM scenario characterized by $(m_{\mbox{\tiny DM}},\sigma_0,\eta_0)$ if oscillations were not present, while the denominator expresses the same quantity when oscillations (with parameter $\delta m$) are  switched-on. $r$ therefore quantifies how much we can reduce the DM density by introducing oscillations. When in the denominator we select the value of $\delta m$ that gives the correct relic abundance $\Omega_{\mbox{\tiny DM}} h^2 \simeq 0.11$, $r$ characterizes therefore the amount of overclosure of the Universe which we would have in the absence of oscillations. We plot $r$ in fig.~\ref{parameterranges}. We work for definiteness in the case with no elastic scatterings ($\xi = 0$), but we do show a line in the case with scatterings (marked $\xi = 10^{-2}$). We also indicate some values in specific points on the iso-lines in fig.~\ref{paramspacezoom}, in this case both for $\xi = 0$ and $\xi = 10^{-2}$. We see that $r$ can reach very large values, i.e. oscillations can be very efficient in depleting the DM excess.

\smallskip

\begin{figure}[!t]
\begin{center}
\includegraphics[width=0.49 \textwidth]{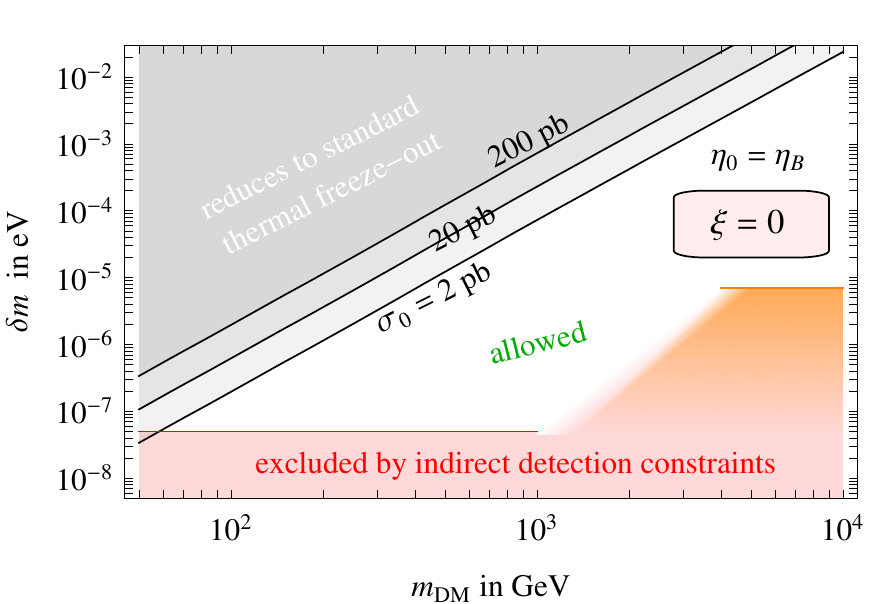}
\includegraphics[width=0.49 \textwidth]{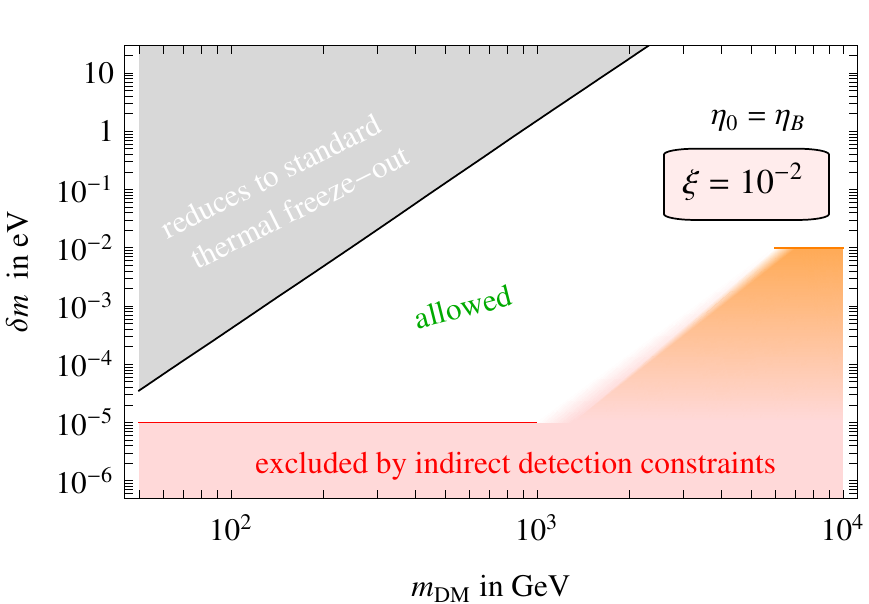}
\caption{\em \small \label{illustrationparamspace} Approximate illustration of the relevant parameter space in the scenarios we are considering, for the case without elastic scatterings ({\bf \em left panel}) and for the case with scatterings ({\bf \em right panel}). The upper left area shaded in grey corresponds to
the region in which the evolution reduces to the standard freeze-out scenario.
The white region refers to a regime in which oscillations recouple annihilations on cosmological scales and we always assume that a suitable value for the annihilation cross section is adopted for a given choice of $m_{\mbox{\tiny \rm DM}}$ and $\delta m$ such that we reproduce the correct relic abundance.
The lower area shaded in pink/orange is excluded by the constraints discussed in Sec.~\ref{sec:constraints}  (red is for {\sc FERMI} and orange for {\sc H.E.S.S.}, see Fig.~\ref{paramspacezoom}) either because it
requires a too high cross section in `our' regime, or because the value of $\delta m$ is such that oscillations do not recouple annihilations on the cosmological scales, and we are back to a usual WIMP scenario.
The fuzzy edge in the large $m_{\mbox{\tiny \rm DM}}$ portion indicates that it is not possible to individuate a single $\delta m$ in the area where the {\sc H.E.S.S.} constraints matter. 
We stress that these figures only illustrate the approximate areas of interest on the basis of eq.~(\ref{deltammaxannscatt}), while the results in all other plots in fig.~\ref{paramspace} and~\ref{paramspacezoom} are determined by the full numerical solutions.}
\end{center}
\end{figure}

One could also wonder whether $\delta m$ can be indefinitely large in these set-ups. This is of course not the case: for too large $\delta m$ 
oscillations start too early and symmetrize the dark sector such that decoupling proceeds as in the standard thermal freeze-out scenario. 
For instance, case E in fig.~\ref{results} illustrates a critical case in which oscillations begin somewhat precisely at the right moment to thwart the impact of the asymmetry and drive the evolution along the usual freeze-out history.
It can therefore also be useful to explicitly define $\delta m _{\rm max}$ as the value of $\delta m$ below which the new phenomena described here arise. The determination of the value of $\delta m _{\rm max}$ which is relevant in the different scenarios is of course tightly related to the identification of the effective start of oscillations $x_{\rm osc}$, which we have discussed in the different cases of Sec.~\ref{sec:formalism}. For the case of oscillations only (neglecting elastic scatterings and annihilations), one obtains
\begin{eqnarray} 
\label{deltammax}
\delta m _{\rm max} & \simeq &2\pi~H(m)/x^2 _{\mbox{\tiny decoupl, asym.}} \nonumber \\
                          &\sim& 10^{-11} \sqrt{g_*} \left( m_{\mbox{\tiny DM}}/1~{\rm GeV}\right) ^2 ~{\rm eV},
\end{eqnarray}
where the numerical estimate in the last step is obtained by neglecting a small change in the value of $x = m_{\mbox{\tiny DM}}/T$ at decoupling, in our scenario with respect to the standard case, i.e. we assumed the standard value $x_{\mbox{\tiny decoupl, asym.}} \approx x_{\mbox{\tiny decoupl, std.}} \sim 20$.
We see, that for  heavy DM, with mass $\sim1$ TeV, already for $\delta m \leq 10^{-5}$ eV oscillations affect the decoupling history. For lighter DM, $\delta m$ is accordingly smaller. 
In the case with annihilations or scatterings, the relation above is modified as discussed in Sec.~\ref{sec:formalism}. One has 
\begin{eqnarray}
\label{deltammaxannscatt}
\delta m _{\rm max} & \simeq & \left( \frac{H_m \, \gamma_{\rm a,s}}{x_{\rm decoupl}} \right)^{1/2} \nonumber \\
& \approx &10^{-7}\, {\rm eV} \sqrt{\frac{g_{*{\rm s}}}{85}\,\frac{\sqrt{g_{*}}}{85}\,\frac{\sigma_0}{1\,{\rm pb}}\,\frac{\eta_0}{\eta_{\mbox{\tiny B}}}\,\left( \frac{m_{\mbox{\tiny DM}}}{100\,{\rm GeV}}\,\frac{20}{x_{\rm decoupl}}\right)^5} \quad {\rm if} \ \gamma_{\rm a} \gg \gamma_{\rm s} \\
& \approx & 4\cdot10^{-4}\, {\rm eV} \sqrt{\left( \frac{g_{*}}{85} \right) ^{3/2}\,\frac{\xi}{10^{-2}}\,\left(\frac{m_{\mbox{\tiny DM}}}{100\,{\rm GeV}}\,\frac{20}{x_{\rm decoupl}}\right)^7} \quad {\rm if}\ \gamma_{\rm s} \gg \gamma_{\rm a}
\end{eqnarray}
where $\gamma_{\rm a,s}$ is the larger of the effective rates introduced in Sec.~\ref{sec:formalism}. 
Figure \ref{illustrationparamspace} illustrates the regions in the parameter plane $\delta m/m_{DM}$ which are individuated by these approximate arguments. 

\smallskip

Another feature of these models worth emphasizing is that the required annihilation cross section always needs to be \emph {higher} than the usual thermal freeze-out value $\sigma_0$. This occurs just because annihilations have to still be active later than in the usual scenario.
The parameter space is indeed effectively bound from below at cross sections of the order of $2$ pb \footnote{Note that as we have two DM species, we have a twice lower number of targets than in the case where DM is its own antiparticle, and therefore cross sections twice higher are needed to have the same annihilation rate.}. High cross sections in the standard case would under-produce $\Omega_{\mbox{\tiny DM}}$, while with the asymmetry+oscillation mechanism, we can reach the correct value.

\medskip

\section{Constraints}
\label{sec:constraints}

\begin{figure}[!t]
\begin{center}
\includegraphics[width=0.49 \textwidth]{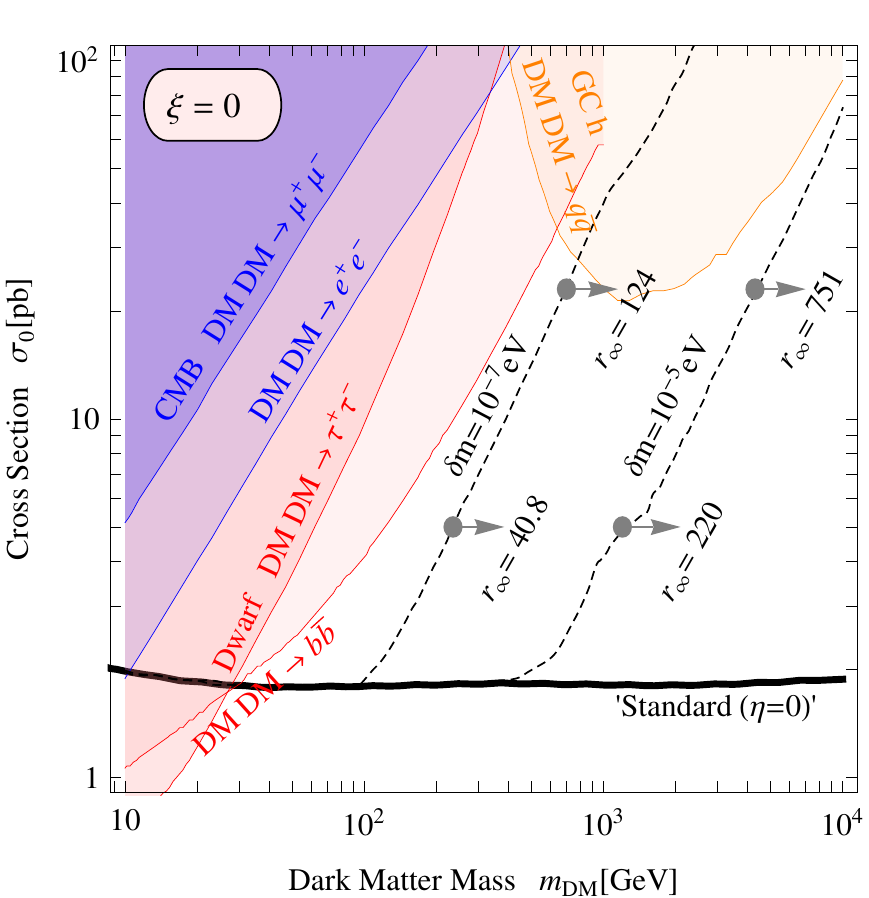}
\includegraphics[width=0.49 \textwidth]{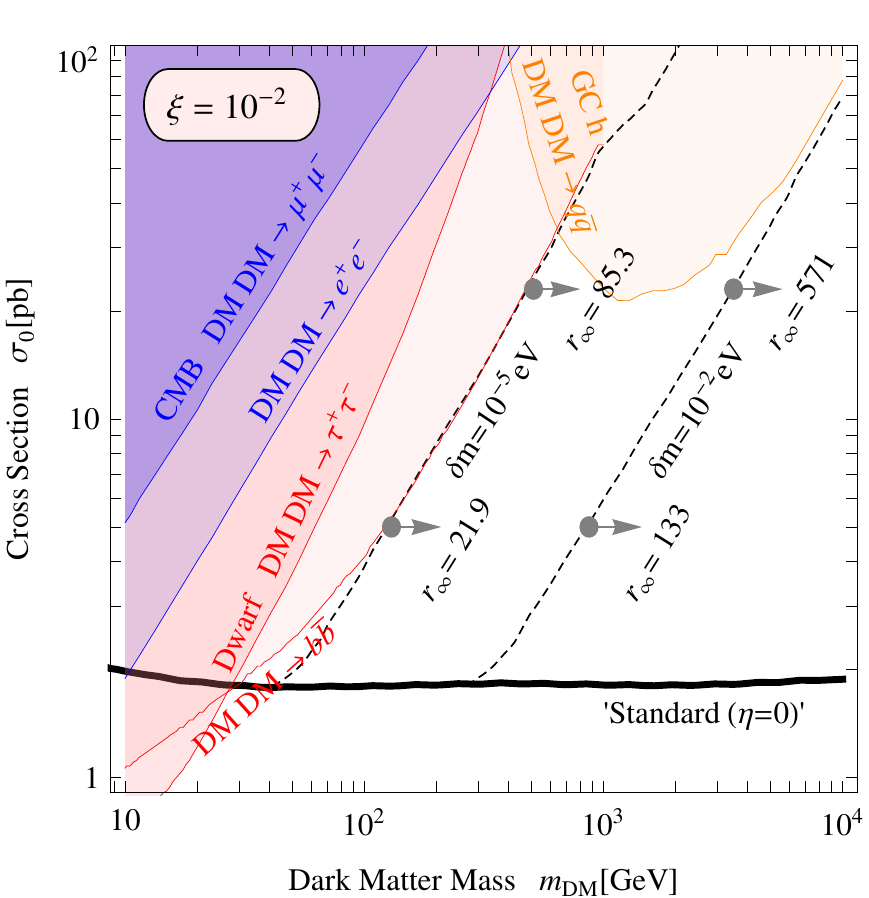}
\caption{\em \small \label{paramspacezoom} Summary plots of the parameter space, showing also the constraints. The dotted lines mark the contours along which a correct $\Omega_{\mbox{\tiny \rm DM}}h^2$ can be obtained. {\bf \em Left}: Oscillations and annihilations only ($\xi = 0$). {\bf \em Right}: Including elastic scatterings ($\xi = 10^{-2}$). In both panels we assume an initial asymmetry $\eta_0 = \eta_{\mbox{\tiny B}}$ and we show two indicative values of the oscillation parameter $\delta m$. The solid  black line at the bottom represents the standard case ($\eta = 0, \delta m = 0$). At some points on the contours, we provide the value of the ratio $r$ defined in eq.~(\ref{eq:ratio}). The shaded blue regions are excluded by CMB constraints, the shaded pink ones are disfavored by gamma ray observations with {\sc FERMI} and the orange ones by observations with {\sc H.E.S.S.} (see text). The white areas above the solid black line are allowed.}
\end{center}
\end{figure}

In the setup we are considering, when oscillations start, annihilations promptly resume. Therefore the parameter space presented above is subject to the usual constraints on annihilating dark matter. Since, in some examples, we are dealing with large annihilation cross sections, these constraints can be particularly significant. We will discuss the constraints coming from the different epochs, and then identify the most stringent ones. 
We will always work under the assumption that oscillations started much before the epoch considered for the constraint, so that the populations $Y^+$ and $Y^-$ have already undergone a very large number of oscillation cycles and therefore can be both approximated with their average value. In this case the annihilation rate is determined only by $\sigma _0$, as usual. In other words, when this condition is satisfied we do not have to worry about the time dependence of the populations of the two species (and therefore of the annihilation rate) or about  possible partial repopulations of one of the two species.

\smallskip

\noindent {\bf BBN.} 
The period of the synthesis of nuclei in the primordial plasma (i.e. Big Bang Nucleosynthesis (BBN)) is the earliest test of standard cosmology, constraining the properties of the Universe starting from when it was a few seconds old, or equivalently at the MeV temperature scale. The good agreement of predicted abundances of the light elements with their measured values makes BBN a powerful cosmological probe: injections of particles and energy due to DM annihilation or decay, either during  BBN, or at later times when those abundances are established, are constrained, as they would alter the observed abundances of primordial elements with respect to prediction (for a review see~\cite{Iocco:2008va}).

More precisely, BBN can offer in principle two types of probes for the scenarios in which we are interested. If oscillations start well before BBN, DM annihilations could be happening at a low level during the BBN (without significantly changing $\Omega_{\mbox{\tiny DM}}$) and the usual constraints on $\sigma _0$ during that era would apply (see e.g. \cite{Hisano:2011dc}). However these constraints are typically weaker than the ones we will discuss below.
A second, more attractive possibility arises if oscillations start {\em after} the end of BBN, i.e. if $t_{\rm osc} > t_{\rm BBN}$. In that case, as annihilations recouple, a large amount of energy is injected into the plasma. The set-up is similar to the one of late-decaying heavy DM progenitor states. Such decays have been extensively studied and stringent constraints set, in the energy injection versus injection time plane. If the characteristic time $t_{\rm osc}$ is longer than 0.1 s, we would in fact be in a position to constrain the amount of energy stored in the dark sector before oscillations start, i.e. the initial value of DM asymmetry $\eta_0$. However, once again, this possibility appears to be ruled out in the set-up in which we are interested, since $t_{\rm osc}\lesssim 0.1$ sec on all the regions of the parameter space which are not already ruled out by the other constraints we discussed below.

\smallskip

\noindent {\bf Epoch of Reionization and CMB.} Strong constraints are imposed on DM annihilations from considering the effect on the generation of the CMB anisotropies at the epoch of recombination (at redshift $\sim1100$) and their subsequent evolution down to the epoch of reionization. The actual physical effect of energy injection around the recombination epoch is that it results in an increased amount of free electrons, which survive to lower redshifts and affect the CMB anisotropies~\cite{CMBconstraints2}. Detailed constraints have been recently derived in~\cite{CMBconstraints}, based on the WMAP (7-year) and Atacama Cosmology Telescope 2008 data. The constraints are somewhat sensitive to the dominant DM annihilation channel: annihilation modes for which a portion of the energy is carried away by neutrinos or stored in protons have a lesser impact on the CMB; on the contrary the annihilation mode which produces directly $e^+e^-$ is the most effective one.

We reproduce in fig.~\ref{paramspacezoom} the constraints on our parameter space as obtained in~\cite{CMBconstraints}, for the two most stringent channels. It happens that the bounds run with the same slope in the ($m_{\mbox{\tiny \rm DM}},$$\sigma_0$) plane as the lines of constant $\Omega_{\mbox{\tiny DM}}$. One sees that, roughly, only the parameter space below $\delta m \geq 10 ^{-9}\, {\rm eV}\, (10^{-7} {\rm eV})$ is allowed for the case of $\xi = 0$ ($\xi = 10^{-2}$), independently on the DM mass. 

\smallskip

\noindent {\bf Present epoch $\mathbf{\gamma}$-rays.} For most of the DM annihilation modes, another relevant constraint is in fact imposed by the indirect DM searches in the present epoch. The DM constraints provided by the {\sc FERMI-LAT} gamma ray data are particularly relevant as they are now cutting into the $\sigma_0 \sim 1$ pb value for low DM mass ($\lesssim 30$ GeV) and a variety of channels.  In particular, dwarf satellite galaxies of the Milky Way are among the most promising targets for dark matter searches in gamma-rays because of their large dynamical mass to light ratio and small expected background from astrophysical sources. No dwarf galaxy has been detected in gamma rays so far and stringent upper limits are placed on DM annihilation by applying a joint likelihood analysis to 10 satellite galaxies with 2 years of {\sc FERMI-LAT} data, taking into account the uncertainty in the dark matter distribution in the satellites~\cite{FERMidwarfs}. The limits are particularly strong for hadronic annihilation channels, and somewhat weaker for leptonic channels as diffusion of leptons out of these systems is poorly constrained. 
On the other hand, strong limits on leptonic ($\mu ^+ \mu ^-$) annihilation channels are set by, for example, the gamma-ray diffuse emission measurement at intermediate latitudes, which probes DM annihilation in our Milky Way halo. In particular, the most recent limits come from 2 years of the {\sc FERMI-LAT} data in the $5^{\circ}\leq b \leq15 ^{\circ},~-80^{\circ}\leq \ell \leq 80^{\circ}$ region~\cite{haloconstraints}, where $b$ and $\ell$ are the galactic latitude and longitude. 

Another relevant constraint in the large mass region is imposed by the observation of the Galactic Center halo with the {\sc H.E.S.S.} telescope~\cite{Abramowski:2011hc}. This refers to a $q \bar q$ annihilation channel and assumes that the DM distribution in the Galaxy follows a Navarro-Frenk-White or Einasto profile (notice that the constraint is lifted in case of a cored profile: the search is made by contrasting the source region closer to the GC with a background region further away, and, in case of a cored profile, both would yield the same DM flux).

We superimpose all these constraints on the plane in fig.~\ref{paramspacezoom}. We see that they are somewhat stronger than the CMB ones considered above. We keep however the latters as they are less model dependent. More generally, we stress that all these constraints are valid under different (somewhat mutually exclusive) assumptions, such as e.g. the DM annihilation channel or the DM galactic profile. As a consequence, while we report all of them on the same plane for completeness, the precise regions of the parameter space which are actually excluded depend on the precise DM model.

\smallskip

\noindent {\bf Collider constraints.} 
Finally, we also mention the constraints imposed by collider searches for dark matter~\cite{collider}: in the framework of an effective field theory, the annihilation cross section of two DM particles into SM particles can be related to their production cross section in proton-antiproton collisions (Tevatron), proton-proton collisions (LHC) or $e^+e^-$ ones (LEP), and therefore constrained by the searches at these respective machines. Current constraints are particularly relevant at low masses, and in some cases compete in strength with the bounds discussed above, while the LHC will soon extend the reach at larger masses. Contrary to the constraints discussed above, however, the collider ones ultimately rely on the assumption that DM couples via contact operators to the SM particles in the initial states of the collision process. E.g. DM annihilating into metastable new light states that then decay into SM particles cannot be constrained this way. We prefer therefore not to show them on the parameter space.

\smallskip

In the light of these bounds, many of the illustrative points presented in Fig.~\ref{results} are excluded since they correspond to low masses, except for point C. 
As our final result, we focus therefore on the allowed region and show in the panels of Fig.~\ref{paramspacezoom}, which are our final summary plots, the contour lines for $\Omega_{\mbox{\tiny DM}} h^2=0.11$. We also indicate, for some chosen points, the value of the ratio $r$ defined in eq.~(\ref{eq:ratio}).


\section{Conclusions}
\label{conclusions}

In this work we have studied the impact of adding oscillations between DM and $\overline{{\rm DM}}$ particles on the scenario of Asymmetric Dark Matter. Such oscillations arise naturally in aDM models, as we argued in Sec.~\ref{sec:theory}, and should therefore be included. 
We found in particular that a typical WIMP with a mass at the EW scale ($\sim$ 100 GeV $-$ 1 TeV) presenting a primordial asymmetry of the same order as the baryon asymmetry naturally gets the correct relic abundance if the $\Delta$(DM) = 2 mass term is in the $\sim$ meV range. This turns out to be a natural value for fermonic DM arising from the higher dimensional operator $H^2\, {\rm DM}^2/\Lambda$ where $H$ is the Higgs field and $\Lambda \sim M_{\rm GUT} - M_{\rm Pl}$.

We have outlined the formalism (based on following the evolution of the density matrix of DM and $\overline{{\rm DM}}$ populations) needed to treat the system of particles that oscillate coherently but at the same time suffer coherence-breaking elastic scatterings on the plasma and annihilations among themselves. Our formalism starts from a standard simplified form of the equation for the density matrix (eq.~(\ref{masterequation})) and  makes use of a few simplifications (see Sec.~\ref{sec:formalism} for a full discussion) but nevertheless is sufficient to illustrate the qualitative features that enter into play and that we want to stress. It would certainly be interesting to write a more rigorous set of equations derived from first principles. This is presently a very active field of research and still under development, in particular in the baryogenesis and leptogenesis community \cite{kadanoffbaym}. Those developements would be relevant for a more precise treatment of the problems that we are interested in here or for further complications of the picture (such as including CP violation in the game).

We have then applied such formalism to explore the phenomenologically available space, by varying the parameters of the dark matter mass $m_{\mbox{\tiny DM}}$, the annihilation cross section $\sigma_0$, the primordial asymmetry in the DM sector $\eta_0$ and the mass difference $\delta m$ which governs oscillations, for two discrete choices of the parameter $\xi$ that sets the strength of the elastic scatterings between DM and the plasma.
The quantitative results are displayed in fig.~\ref{paramspace}, which illustrates the exploration of the parameter space, and in fig.~\ref{paramspacezoom}, which takes into account the constraints and focuses on the still allowed regions.

Our main result is readily summarized: we have shown that the parameter space at disposal becomes much wider and richer and in particular it is possible to have models of asymmetric DM with large $m_{\mbox{\tiny DM}}$ (and large annihilation cross section) while still reproducing the right relic abundance $\Omega_{\mbox{\tiny DM}} h^2$ in the Universe. In other words, the mass and annihilation cross section of asymmetric WIMP dark matter are almost unconstrained by the relic density condition (for masses above $\sim 10$ GeV and for annihilation cross sections larger than the thermal one), as soon as we introduce oscillations. 
The main physical reason and the underlying mechanism are simple to understand: while in aDM the population of DM particles is frozen by the lack of targets, the oscillations re-symmetrize DM and $\overline{{\rm DM}}$ particles, therefore allowing annihilations (if strong enough) to deplete them further until freeze-out; a smaller final number density accommodates a larger $m_{\mbox{\tiny DM}}$.

\begin{table}[t]
\begin{center}
\begin{tabular}{|c| c c|}
\hline
&  &\\
symmetric DM& $\Omega_{\mbox{\tiny DM}} \propto {\sigma_0^{-1}}$ &  \\ 
&     &  \\ \hline
    &  &\\ 

& $\bullet$ \underline{if $\xi=0$ (negligible scattering)}: &\\[2mm] 
       
&  $\Omega_{\mbox{\tiny DM}} \simeq \frac{\displaystyle{m_{\mbox{\tiny DM}} \, s}}{\displaystyle{\rho_{\rm crit}}} \, \eta_0  \ {\Bigg[1+     4806 \,  
\left( \displaystyle \frac{g_{* {\rm s},\infty}^4}{g_{*,\infty}^3} \
\frac{\delta m^2}{{\rm eV}^2} 
\left( \frac{\sigma_0}{{\rm pb}} \right)^4
\left(\frac{\eta_0}{\eta_{\mbox{\tiny B}}} \right)^4   
	\right)^{1/5}  \Bigg]^{-1}} $ &   \\[7mm]

&  $ \Omega_{\mbox{\tiny DM}}  \approx   
\frac{\displaystyle{m_{\mbox{\tiny DM}} \, s}}{\displaystyle{\rho_{\rm crit}}} \  
\left\{ \begin{array}{ll}
	\eta_0 &  \hbox{if} \ \delta m \ll \delta m^\prime \\

	\displaystyle 2 \cdot 10^{-12} \ \eta_0^{1/5} 
	\left( \frac{g_{*,\infty}^3}{g_{*{\rm s},\infty}^4} \,
		\frac{{\rm eV}^2}{\delta m^2} \,
		\frac{{\rm pb}^4}{\sigma_0^4} 
	\right)^{1/5}  &  			\hbox{if} \ \delta m \gg \delta m^\prime  \\
	
	\end{array} \right. $ & \\[9mm]
   
&  \hbox{with} $\delta m^{\prime} \approx  6.2 \cdot 10^{-10} \ \frac{g_{*,\infty}^{3/2}}{g_{*{\rm s},\infty}^2} \left( \frac{{\rm pb}}{\sigma_0} \right)^2  \left(\frac{\eta_{\mbox{\tiny B}}}{\eta_0} \right)^2 $ eV &\\ 
asymmetric DM  &  &\\ \cline{2-3}
   
   & & \\
   
& $\bullet$ \underline{if $\xi \neq0$ (scattering with primordial plasma)}: &\\[2mm] 

   &  $\Omega_{\mbox{\tiny DM}} \simeq \frac{\displaystyle{m_{\mbox{\tiny DM}} \, s}}{\displaystyle{\rho_{\rm crit}}} \, \eta_0 \ 
              
       {\Bigg[1+  6.94 \, \displaystyle \frac{g_{* {\rm s},\infty}}{g_{*,\infty}^{4/7}} \left(\frac{\delta m /{\rm eV}}{\xi} \right)^{2/7} \frac{\sigma_0}{1 \, {\rm pb}} \, \frac{\eta_0}{\eta_{\mbox{\tiny B}}} \Bigg]^{-1}}$ &   \\[7mm]      
              
&  $ \Omega_{\mbox{\tiny DM}}  \approx  \frac{\displaystyle{m_{\mbox{\tiny DM}} \, s}}{\displaystyle{\rho_{\rm crit}}} \  \left\{ \begin{array}{ll}
 \eta_0 &   \hbox{if} \ \delta m \ll \delta m^{\prime\prime} \\

1.4 \cdot 10^{-11} \,\xi \, \displaystyle \frac{g_{*,\infty}^{4/7}}{g_{* {\rm s},\infty}} \frac{{\rm pb}}{\sigma_0}  \left(\frac{{\rm eV}}{\delta m} \right)^{2/7}  &  \hbox{if} \ \delta m \gg \delta m^{\prime\prime}\\
	\end{array} \right. $ & \\[9mm]
 
 &  \hbox{with} $\delta m^{\prime\prime} \approx 10^{-3} \, \xi \left( \frac{g_{* {\rm s},\infty}}{g_{*,\infty}^{4/7}}  \frac{\sigma_0}{1 \, {\rm pb}} \frac{\eta_0}{\eta_{\mbox{\tiny B}}} \right)^{-7/2} $ eV &\\[5mm] 
\hline
\end{tabular}
\end{center}
\caption{
\label{tablesummary}
\small Scaling of the WIMP relic abundance with the DM parameters $m_{\mbox{\tiny DM}}$, $\delta m$, $\sigma_0$ and  the primordial asymmetry $\eta_0$. The scattering of DM with the plasma is parametrized by  $\xi=G_{\mbox{\tiny DM}}/G_{\rm F} $ where $G_{\mbox{\tiny DM}}$ is the analog of the Fermi constant for the coupling of DM to matter. The formul\ae\ follow eq.~(\ref{anapproximationA})$-$(\ref{anapproximationB}) and are valid for $\sigma_0$ not too large (see text in Sec.~\ref{sec:results} for details).
}
\end{table}

In this setup, therefore, $\Omega_{\mbox{\tiny DM}} h^2$ is no longer solely controlled by the primordial asymmetry $\eta_0$, like in the aDM case, and no longer solely controlled by the annihilation cross section $\langle \sigma v \rangle$, like in the thermal freeze-out case, but instead a smooth bridge is provided between the two. The scaling of $\Omega_{\mbox{\tiny DM}}$ with parameters is given in eq.~(\ref{anapproximationA})$-$(\ref{anapproximationB}) and it is also summarized explicitly in Table \ref{tablesummary}.
These scenarios can thus preserve the attractive feature of aDM, that relates the DM primordial asymmetry and the baryon asymmetry in the first place, but at the same time preserve also the appeal of weak-scale DM mass (and possibly cross-sections). Note that vanilla aDM models with $m_{\mbox{\tiny \rm DM}} \sim$ a few GeV are ruled out unless $\delta m$ is effectively zero.

As a last remark, we note that the dark sector may turn out to be more complicated than studied here. For instance, the WIMP sector may contain different flavours that could mix or there could be mixing between the WIMP and standard neutrinos. These effects have been extensively discussed in the case of sterile neutrino dark matter but not so in the case of WIMP dark matter, apart from the case of sneutrinos, e.g \cite{MarchRussell:2009aq}.
A recent interest for these possibilities has arisen \cite{Falkowski:2011xh,Cui:2011qe,Agrawal:2011ze}, although the consequences for the relic abundance of WIMPs have not yet been studied in detail (see however the recent Ref.~\cite{Baer:2011uz}). A new future interesting direction of investigation is therefore opening.

\begin{figure}[!h]
\begin{center}
\includegraphics[width=0.49 \textwidth]{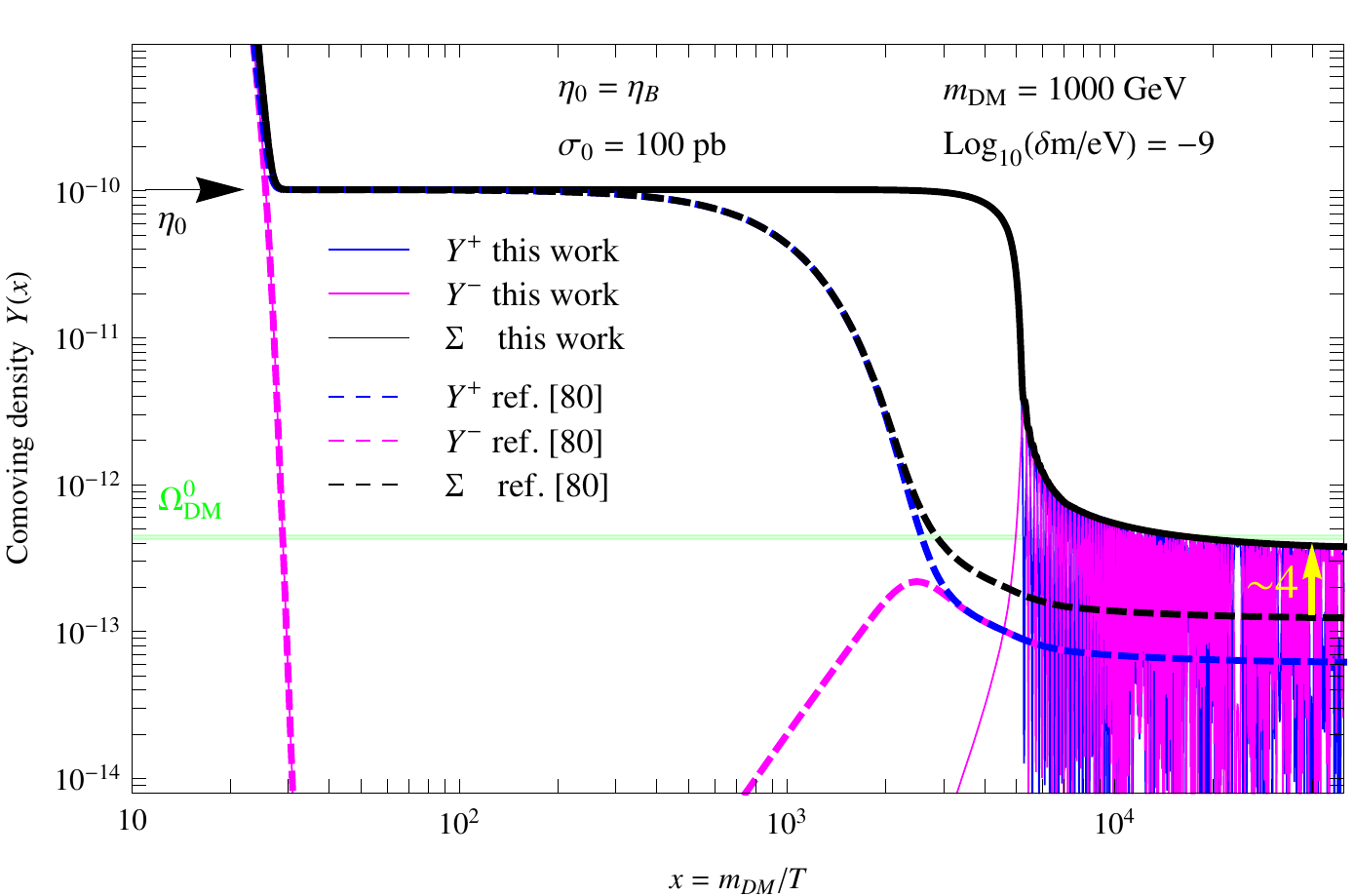}
\includegraphics[width=0.49 \textwidth]{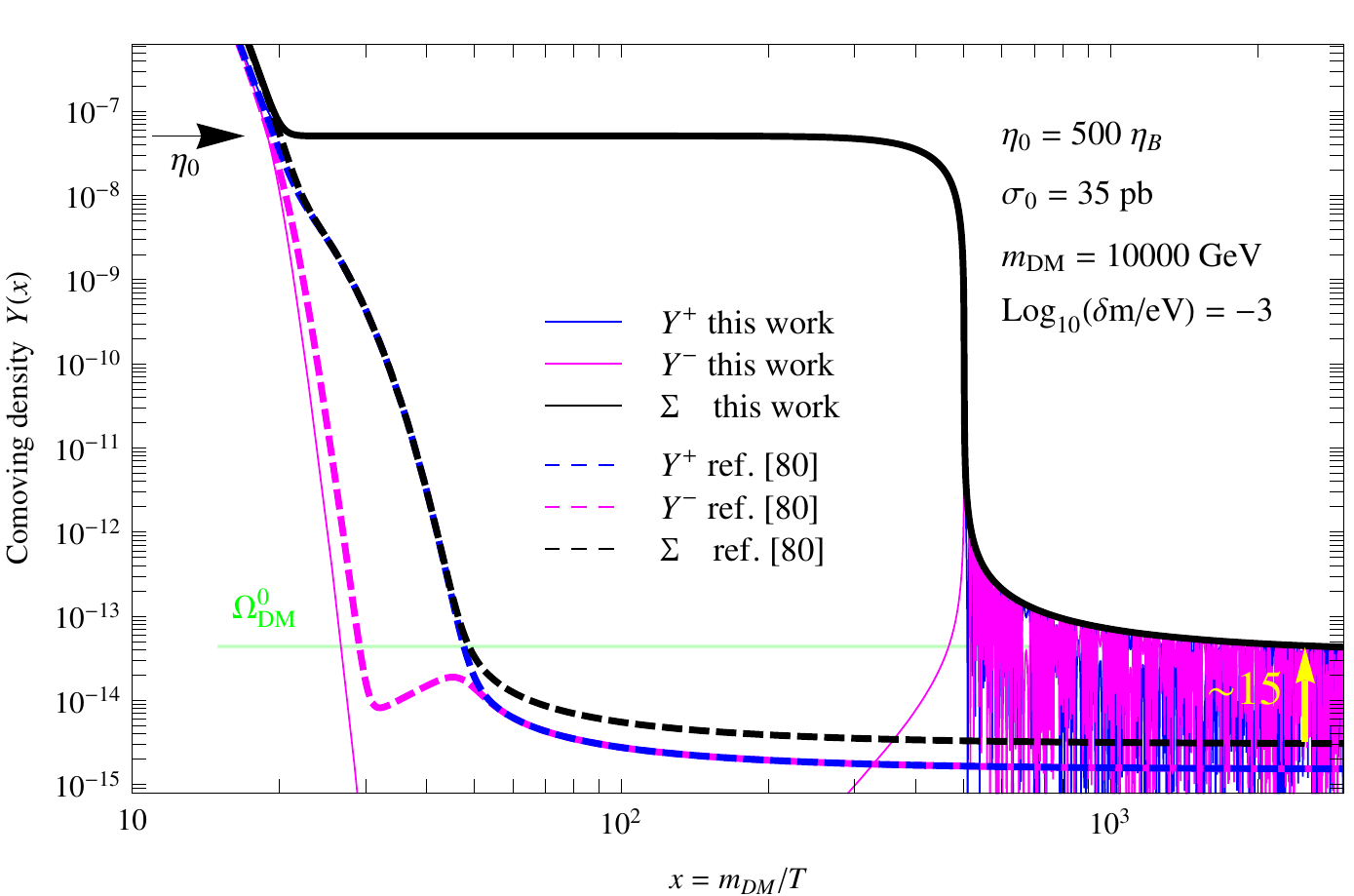}
\caption{\em \small \label{comparisonPB} Comparison with the results of~\cite{Buckley:2011ye}, for two specific choices of parameters (indicated).}
\end{center}
\end{figure}

\paragraph{Note.} While this work was being completed, Ref.~\cite{Buckley:2011ye}, whose scope is similar to ours, appeared. Contrasting our procedure with theirs, we note that they used a simplified Boltzmann equation that treats the oscillations with a constant rate and in a way that does not include decoherence effects due to annihilations. They also do not include elastic scatterings.  
A comparison between the formalisms can be readily done using the Boltzmann-like equations (\ref{Beqannosc}) derived from the matrix formalism: in our full form, the differential equation for the sum $\Sigma=Y^+ + Y^-$ is affected by the evolution of the difference $\Delta=Y^+ - Y^-$ as well as the evolution of the off-diagonal elements of  the density matrix $\Xi$. In Ref.~\cite{Buckley:2011ye}, $\Xi'(x)=0$ and the equation for $\Delta$ does not depend on annihilations nor scatterings, namely $\Delta'(x)/\Delta =-2 \, \delta m /(x H(x))$, in contrast with what Eq.~(\ref{Beqannosc}) indicates (see the discussion at page \pageref{discussion on deltam2/gamma}). This leads to a different damping factor in the evolution of $\Sigma(x)$. 
The simplified approach of~\cite{Buckley:2011ye} can lead to a different or even very different time evolution with respect to the full approach that we pursue. Quantitatively, the discrepancy between the simplified and the full approaches varies depending on the choices of parameters. 
For a significant range of parameter space, both approaches lead to the same relic abundance (up to $\mathcal{O}(1)$ deviations), as we illustrate in an example in Fig.~\ref{comparisonPB} on the left. For other choices, however, the two approaches produce a final relic abundance that differs by more than one order of magnitude (see e.g. the right panel in Fig.~\ref{comparisonPB}).  
To recap, our formalism has the advantage of allowing to describe oscillations consistently as well as to take into account elastic scatterings. Finally, we have provided a full exploration of the parameter space and we are interested in identifying the choices of parameters for which the correct $\Omega_{\mbox{\tiny DM}}$ is obtained, while~\cite{Buckley:2011ye} focusses on two specific values of the DM mass and does not require that the correct DM relic abundance is reproduced.

\paragraph{Acknowledgements}
We thank Zurab Berezhiani, Adam Falkowski, Alex Friedland, Tho-mas Konstandin, Kimmo Kainulainen, Thomas Schwetz-Mangold, Thomas Hambye and Pasquale Serpico for useful discussions. This work is supported by the ERC starting Grant Cosmo@LHC, by the French national research agency ANR under contract ANR 2010 BLANC 041301 and by the EU ITN network UNILHC.  Part of it was completed during the CERN TH-Institute {\it DMUH'11} (18-29 july 2011). 
We also acknowledge support  from the Institut de Physique of CNRS within the `{\it PEPS 2010 Projets - Physique Th\'eorique et ses Interfaces}'.

\bigskip
\bigskip

\footnotesize
\begin{multicols}{2}
  
\end{multicols}


\begin{thebibliography}{nn}

\bibitem{cosmoDM}
See e.g. E.~Komatsu {\it et al.} [ WMAP Collaboration ],
  Astrophys.\ J.\ Suppl.\  {\bf 192 } (2011)  18
  [arXiv:\hhref{1001.4538v3} [astro-ph.CO]].  



\bibitem{Gelmini:2010zh}
See e.g.  G.~Gelmini, P.~Gondolo,
  In {\it Bertone, G. (ed.): Particle Dark Matter}, 121-141 
  [arXiv:\hhref{1009.3690} [astro-ph.CO]];
  B.~S.~Acharya, G.~Kane, S.~Watson and P.~Kumar,
  Phys.\ Rev.\  D {\bf 80} (2009) 083529
  [arXiv:\hhref{0908.2430} [astro-ph.CO]];
  D.~J.~H.~Chung, E.~W.~Kolb, A.~Riotto,
  Phys.\ Rev.\ Lett.\  {\bf 81 } (1998)  4048-4051 
  [\hhref{hep-ph/9805473}].

\bibitem{Boyarsky:2009ix}
  A.~Boyarsky, O.~Ruchayskiy, M.~Shaposhnikov,
  Ann.\ Rev.\ Nucl.\ Part.\ Sci.\  {\bf 59 } (2009)  191-214 
  [arXiv:\hhref{0901.0011} [hep-ph]].

\bibitem{Hall:2009bx}
  L.~J.~Hall, K.~Jedamzik, J.~March-Russell, S.~M.~West,
  JHEP {\bf 1003}, 080 (2010) 
  [arXiv:\hhref{0911.1120} [hep-ph]].


\bibitem{Nussinov:1985xr}
  S.~Nussinov,
  Phys.\ Lett.\  B {\bf 165} (1985) 55.

\bibitem{Barr:1990ca}
  S.~M.~Barr, R.~S.~Chivukula and E.~Farhi,
  Phys.\ Lett.\  B {\bf 241} (1990) 387.

\bibitem{Barr:1991qn}
  S.~M.~Barr,
  Phys.\ Rev.\  {\bf D44 } (1991)  3062-3066.
  
\bibitem{Kaplan:1991ah}
  D.~B.~Kaplan,
  Phys.\ Rev.\ Lett.\  {\bf 68} (1992) 741.

\bibitem{GudnasonKouvarisSannino}
  S.~B.~Gudnason, C.~Kouvaris, F.~Sannino,
  Phys.\ Rev.\  {\bf D73 } (2006)  115003 
  [\hhref{hep-ph/0603014}].  
  
  S.~B.~Gudnason, C.~Kouvaris, F.~Sannino,
  Phys.\ Rev.\  {\bf D74 } (2006)  095008 
  [\hhref{hep-ph/0608055}].

\bibitem{mirror1}
  H.~M.~Hodges,
  Phys.\ Rev.\  {\bf D47 } (1993)  456-459.

\bibitem{mirror2}
R.~Foot, R.~R.~Volkas,
  Phys.\ Rev.\  {\bf D52 } (1995)  6595-6606 
  [\hhref{hep-ph/9505359}].

\bibitem{mirror3}
Z.~G.~Berezhiani, R.~N.~Mohapatra,
  Phys.\ Rev.\  {\bf D52 } (1995)  6607-6611 
  [\hhref{hep-ph/9505385}].
  
\bibitem{mirror4}
Z.~G.~Berezhiani, A.~D.~Dolgov, R.~N.~Mohapatra,
  Phys.\ Lett.\  {\bf B375 } (1996)  26-36 
  [\hhref{hep-ph/9511221}].

\bibitem{mirror5}
L.~Bento, Z.~Berezhiani,
  [\hhref{hep-ph/0111116}].

\bibitem{mirror6}
Z.~Berezhiani,
  ``Through the looking-glass: Alice's adventures in mirror world,''
  In {\it Shifman, M. (ed.) et al.: From fields to strings, vol. 3}, 2147-2195 
  [\hhref{hep-ph/0508233}].

\bibitem{mirror7}
   R.~Foot, R.~R.~Volkas,
  Phys.\ Rev.\  {\bf D68 } (2003)  021304.
  [\hhref{hep-ph/0304261}]. 
  R.~Foot, R.~R.~Volkas,
  Phys.\ Rev.\  {\bf D69 } (2004)  123510.
  [\hhref{hep-ph/0402267}].
  
\bibitem{Farrar:2005zd}
  G.~R.~Farrar, G.~Zaharijas,
  Phys.\ Rev.\ Lett.\  {\bf 96 } (2006)  041302 
  [\hhref{hep-ph/0510079}].

\bibitem{Hooper:2004dc}
  D.~Hooper, J.~March-Russell, S.~M.~West,
  Phys.\ Lett.\  {\bf B605 } (2005)  228-236 
  [\hhref{hep-ph/0410114}].

\bibitem{Kitano:2004sv}
  R.~Kitano, I.~Low,
  Phys.\ Rev.\  {\bf D71 } (2005)  023510 
  [\hhref{hep-ph/0411133}].

\bibitem{Agashe:2004bm}
  K.~Agashe, G.~Servant,
  JCAP {\bf 0502 } (2005)  002 
  \hhref{hep-ph/0411254}].

\bibitem{Cosme:2005sb}
  N.~Cosme, L.~Lopez Honorez, M.~H.~G.~Tytgat,
  Phys.\ Rev.\  {\bf D72 } (2005)  043505 
  [\hhref{hep-ph/0506320}].

\bibitem{Khlopov:2005ew}
K.~Belotsky, D.~Fargion, M.~Khlopov, R.~V.~Konoplich,
  Phys.\ Atom.\ Nucl.\  {\bf 71 } (2008)  147-161.
  [\hhref{hep-ph/0411093}].
M.~Y.~Khlopov,
  Pisma Zh.\ Eksp.\ Teor.\ Fiz.\  {\bf 83 } (2006)  3-6.
  [\hhref{astro-ph/0511796}].

  
  \bibitem{Kaplan:2009ag}
  D.~E.~Kaplan, M.~A.~Luty and K.~M.~Zurek,
  Phys.\ Rev.\  D {\bf 79} (2009) 115016
  [arXiv: \hhref{0901.4117} [hep-ph]].
  
  \bibitem{Cohen:2009fz}
  T.~Cohen and K.~M.~Zurek,
  Phys.\ Rev.\ Lett.\  {\bf 104} (2010) 101301
  [arXiv:\hhref{0909.2035} [hep-ph]].
  
\bibitem{Cai:2009ia}
  Y.~Cai, M.~A.~Luty, D.~E.~Kaplan,
  [arXiv: \hhref{0909.5499} [hep-ph]].

\bibitem{An:2009vq}
  H.~An, S.-L.~Chen, R.~Mohapatra, Y.~Zhang,
  JHEP {\bf 1003 } (2010) 124
  [arXiv:\hhref{0911.4463} [hep-ph]].

  \bibitem{Shelton:2010ta}
  J.~Shelton and K.~M.~Zurek,
  Phys.\ Rev.\  D {\bf 82} (2010) 123512
  [arXiv:\hhref{1008.1997} [hep-ph]].

\bibitem{Buckley:2010ui}
  M.~R.~Buckley, L.~Randall,
  JHEP {\bf 1109 } (2011)  009.
  [arXiv:\hhref{1009.0270} [hep-ph]].
  
  \bibitem{Davoudiasl:2010am}
  H.~Davoudiasl, D.~E.~Morrissey, K.~Sigurdson and S.~Tulin,
  Phys.\ Rev.\ Lett.\  {\bf 105} (2010) 211304
  [arXiv:\hhref{1008.2399} [hep-ph]].
  
\bibitem{Haba:2010bm}
  N.~Haba, S.~Matsumoto,
  Prog.\ Theor.\ Phys.\  {\bf 125 } (2011)  1311-1316.
  [arXiv:\hhref{1008.2487} [hep-ph]].

  \bibitem{Chun:2010hz}
  E.~J.~Chun,
  Phys.\ Rev.\  D {\bf 83} (2011) 053004
  [arXiv:\hhref{1009.0983} [hep-ph]].
      
\bibitem{Gu:2010ft}
  P.-H.~Gu, M.~Lindner, U.~Sarkar, X.~Zhang,
  Phys.\ Rev.\  {\bf D83 } (2011)  055008.
  [arXiv:\hhref{1009.2690} [hep-ph]].
  
\bibitem{Blennow:2010qp}
  M.~Blennow, B.~Dasgupta, E.~Fernandez-Martinez, N.~Rius,
  JHEP {\bf 1103 } (2011)  014.
  [arXiv:\hhref{1009.3159} [hep-ph]].
  
  \bibitem{McDonald:2010rn}
  J.~McDonald,
  arXiv:\hhref{1009.3227} [hep-ph].
  
  \bibitem{Allahverdi:2010rh}
  R.~Allahverdi, B.~Dutta, K.~Sinha,
  Phys.\ Rev.\  {\bf D83 } (2011)  083502.
  [arXiv:\hhref{1011.1286} [hep-ph]].
  
  \bibitem{Dutta:2010va}
  B.~Dutta, J.~Kumar,
  Phys.\ Lett.\  {\bf B699 } (2011)  364-367.
  [arXiv:\hhref{1012.1341} [hep-ph]].

  \bibitem{Falkowski:2011xh}
   A.~Falkowski, J.~T.~Ruderman, T.~Volansky,
  JHEP {\bf 1105 } (2011)  106.
  [arXiv:\hhref{1101.4936} [hep-ph]].

\bibitem{Cheung:2011if}
  C.~Cheung, K.~M.~Zurek,
  Phys.\ Rev.\  {\bf D84 } (2011)  035007.
  [arXiv:\hhref{1105.4612} [hep-ph]].
  
  \bibitem{DelNobile:2011je}
  E.~Del Nobile, C.~Kouvaris, F.~Sannino,
  Phys.\ Rev.\  {\bf D84 } (2011)  027301.
  [arXiv:\hhref{1105.5431} [hep-ph]].

\bibitem{Cui:2011qe}
Y.~Cui, L.~Randall, B.~Shuve,
  [arXiv:\hhref{1106.4834} [hep-ph]].
  
\bibitem{MarchRussell:2011fi}
  J.~March-Russell, M.~McCullough,
  [arXiv:\hhref{1106.4319} [hep-ph]].

\bibitem{Frandsen:2011kt}
  M.~T.~Frandsen, S.~Sarkar, K.~Schmidt-Hoberg,
  Phys.\ Rev.\  {\bf D84 } (2011)  051703.
  [arXiv:\hhref{1103.4350} [hep-ph]].

\bibitem{Buckley:2011kk}
  M.~R.~Buckley,
  Phys.\ Rev.\  {\bf D84 } (2011)  043510.
  [arXiv:\hhref{1104.1429} [hep-ph]].

\bibitem{Davoudiasl:2011fj}
  H.~Davoudiasl, D.~E.~Morrissey, K.~Sigurdson, S.~Tulin,
  [arXiv:\hhref{1106.4320} [hep-ph]].

\bibitem{Graesser:2011vj}
  M.~L.~Graesser, I.~M.~Shoemaker, L.~Vecchi,
  [arXiv:\hhref{1107.2666} [hep-ph]].

\bibitem{Arina:2011cu}
  C.~Arina, N.~Sahu,
  [arXiv:\hhref{1108.3967} [hep-ph]].

\bibitem{McDonald:2011sv}
  J.~McDonald,
    [arXiv:\hhref{1108.4653} [hep-ph]].

\bibitem{Barr:2011cz}
  S.~M.~Barr,
  [arXiv:\hhref{1109.2562} [hep-ph]].


\bibitem{direct}
  A.~L.~Fitzpatrick, D.~Hooper, K.~M.~Zurek,
  Phys.\ Rev.\  {\bf D81 } (2010)  115005 
  [arXiv:\hhref{1003.0014} [hep-ph]].
    H.~An, S.~-L.~Chen, R.~N.~Mohapatra, S.~Nussinov, Y.~Zhang,
  Phys.\ Rev.\  {\bf D82 } (2010)  023533 
  [arXiv:\hhref{1004.3296} [hep-ph]].
   N.~Fornengo, P.~Panci, M.~Regis,
  [arXiv:\hhref{1108.4661} [hep-ph]].


     
\bibitem{Graesser:2011wi}
  M.~L.~Graesser, I.~M.~Shoemaker and L.~Vecchi,
  arXiv:\hhref{1103.2771} [hep-ph].
  
\bibitem{Iminniyaz:2011yp}
  H.~Iminniyaz, M.~Drees and X.~Chen,
  arXiv:\hhref{1104.5548} [hep-ph].
    


  
    
  \bibitem{bounds}
  M.~T.~Frandsen and S.~Sarkar,
  Phys.\ Rev.\ Lett.\  {\bf 105} (2010) 011301
  [arXiv:\hhref{1003.4505} [hep-ph]].
  B.~Feldstein, A.~L.~Fitzpatrick,
  JCAP {\bf 1009}, 005 (2010).
  [arXiv:\hhref{1003.5662} [hep-ph]].
  D.~T.~Cumberbatch, J.~.A.~Guzik, J.~Silk, L.~S.~Watson, S.~M.~West,
  Phys.\ Rev.\  {\bf D82 } (2010)  103503 
  [arXiv:\hhref{1005.5102} [astro-ph.SR]].
  M.~Taoso, F.~Iocco, G.~Meynet, G.~Bertone and P.~Eggenberger,
  Phys.\ Rev.\  D {\bf 82} (2010) 083509
  [arXiv:\hhref{1005.5711} [astro-ph.CO]].
  C.~Kouvaris, P.~Tinyakov,
  Phys.\ Rev.\  {\bf D83 } (2011)  083512 
  [arXiv:\hhref{1012.2039} [astro-ph.HE]].
  S.~D.~McDermott, H.~-B.~Yu, K.~M.~Zurek,
   [arXiv:\hhref{1103.5472} [hep-ph]].
  C.~Kouvaris and P.~Tinyakov,
  arXiv:\hhref{1104.0382} [astro-ph.CO].
       
       
       
\bibitem{Bilenky:1987ty}
  S.~M.~Bilenky and S.~T.~Petcov,
  Rev.\ Mod.\ Phys.\  {\bf 59} (1987) 671
   [Erratum-ibid.\  {\bf 61} (1989) 169]
   [Erratum-ibid.\  {\bf 60} (1988) 575].
       
       
\bibitem{Hirsch:1997vz}
  M.~Hirsch, H.~V.~Klapdor-Kleingrothaus, S.~G.~Kovalenko,
  Phys.\ Lett.\  {\bf B398 } (1997)  311-314.
  [\hhref{hep-ph/9701253}].

\bibitem{Grossman:1997is}
  Y.~Grossman, H.~E.~Haber,
  Phys.\ Rev.\ Lett.\  {\bf 78 } (1997)  3438-3441.
  [\hhref{hep-ph/9702421}].

\bibitem{Hall:1997ah}
  L.~J.~Hall, T.~Moroi, H.~Murayama,
  Phys.\ Lett.\  {\bf B424 } (1998)  305-312.
  [\hhref{hep-ph/9712515}].

\bibitem{Choi:2001fka}
  K.~Choi, K.~Hwang, W.~Y.~Song,
  Phys.\ Rev.\ Lett.\  {\bf 88 } (2002)  141801.
  [\hhref{hep-ph/0108028}].


 
\bibitem{TuckerSmith:2001hy}
  D.~Tucker-Smith, N.~Weiner,
  Phys.\ Rev.\  {\bf D64 } (2001)  043502.
  [\hhref{hep-ph/0101138}].

\bibitem{Cui:2009xq}
  Y.~Cui, D.~E.~Morrissey, D.~Poland, L.~Randall,
  JHEP {\bf 0905 } (2009)  076.
  [arXiv:\hhref{0901.0557} [hep-ph]].


\bibitem{Lindner:2011it}
  M.~Lindner, D.~Schmidt, T.~Schwetz,
  [arXiv:\hhref{1105.4626} [hep-ph]].
See also:  E.~J.~Chun,
  Phys.\ Lett.\  B {\bf 525} (2002) 114
  [arXiv:\hhref{hep-ph/0105157}].

\bibitem{Ma:2006km}
  E.~Ma,
  Phys.\ Rev.\  {\bf D73 } (2006)  077301.
  [\hhref{hep-ph/0601225}].

       
 
\bibitem{formalism}    
The formalism for the propagation of interacting and oscillating neutrinos has been fully presented in 
G.~Raffelt, G.~Sigl, L.~Stodolsky,
  Phys.\ Rev.\ Lett.\  {\bf 70 } (1993)  2363-2366 
  [\hhref{hep-ph/9209276}], 
G.~Sigl, G.~Raffelt,
  Nucl.\ Phys.\  {\bf B406 } (1993)  423-451, 
  although it had been pioneeringly introduced in 
A.~D.~Dolgov,
  Sov.\ J.\ Nucl.\ Phys.\  {\bf 33 } (1981)  700-706 (A.~D.~Dolgov, Yad.\ Fiz.\ {\bf 33}, 1309 (1981) 
and 
R.~Barbieri, A.~Dolgov,
  Nucl.\ Phys.\  {\bf B349 } (1991)  743-753.
A collection of results is found in 
A.~D.~Dolgov,
  Phys.\ Rept.\  {\bf 370 } (2002)  333-535 
  [\hhref{hep-ph/0202122}].
    
\bibitem{ScherrerTurner}
See 
  R.~J.~Scherrer, M.~S.~Turner,
  Phys.\ Rev.\  {\bf D34 } (1986)  3263,
which is the {\it Erratum} of the Appendix of  
  R.~J.~Scherrer, M.~S.~Turner,
  Phys.\ Rev.\  {\bf D33 } (1986)  1585.
    
\bibitem{neutrinoreviews}    
M.~C.~Gonzalez-Garcia, Y.~Nir,
  Rev.\ Mod.\ Phys.\  {\bf 75 } (2003)  345-402 
  [\hhref{hep-ph/0202058}].
A.~Strumia, F.~Vissani,
  [\hhref{hep-ph/0606054}].  
  
  
\bibitem{Abazajian:2001nj}
  K.~Abazajian, G.~M.~Fuller, M.~Patel,
  Phys.\ Rev.\  {\bf D64}, 023501 (2001).
  [astro-ph/0101524].


\bibitem{Dodelson:1993je}
  S.~Dodelson, L.~M.~Widrow,
  Phys.\ Rev.\ Lett.\  {\bf 72}, 17-20 (1994).
  [hep-ph/9303287].  
    

\bibitem{Iocco:2008va}
  F.~Iocco, G.~Mangano, G.~Miele, O.~Pisanti, P.~D.~Serpico, 
  Phys.\ Rept.\  {\bf 472}, 1-76 (2009) 
  [arXiv:\hhref{0809.0631} [astro-ph]].

\bibitem{Hisano:2011dc}
  J.~Hisano, M.~Kawasaki, K.~Kohri, T.~Moroi, K.~Nakayama, T.~Sekiguchi,
  Phys.\ Rev.\  {\bf D83}, 123511 (2011) 
  [arXiv:\hhref{1102.4658} [hep-ph]].

\bibitem{CMBconstraints2}
  S.~Galli, F.~Iocco, G.~Bertone, A.~Melchiorri,
  Phys.\ Rev.\  {\bf D80 } (2009)  023505 
  [arXiv:\hhref{0905.0003} [astro-ph.CO]].
  
  T.~R.~Slatyer, N.~Padmanabhan, D.~P.~Finkbeiner,
  Phys.\ Rev.\  {\bf D80 } (2009)  043526  
  [arXiv:\hhref{0906.1197} [astro-ph.CO]].
  
  G.~Huetsi, A.~Hektor, M.~Raidal,
  Astron.\ Astrophys.\  {\bf 505 } (2009)  999-1005 
  [arXiv:\hhref{0906.4550} [astro-ph.CO]].
  
  M.~Cirelli, F.~Iocco, P.~Panci,
  JCAP {\bf 0910 } (2009)  009 
  [arXiv:\hhref{0907.0719} [astro-ph.CO]].

\bibitem{CMBconstraints}
G.~Hutsi, J.~Chluba, A.~Hektor, M.~Raidal 
  [arXiv:\hhref{1103.2766} [astro-ph.CO]].

S.~Galli, F.~Iocco, G.~Bertone, A.~Melchiorri 
  [arXiv:\hhref{1106.1528} [astro-ph.CO]].

\bibitem{FERMidwarfs}
The {\sc FERMI-LAT} collaboration,
  [arXiv: \hhref{1108.3546} [astro-ph.HE]].
See also: talk by Maja L. Garde at the \myurl{fermi.gsfc.nasa.gov/science/symposium/2011/program/}{FERMI Symposium 2011}.

\bibitem{haloconstraints}
  M.~Cirelli, P.~Panci, P.~D.~Serpico,
  Nucl.\ Phys.\  {\bf B840}, 284-303 (2010) 
  [arXiv:\hhref{0912.0663} [astro-ph.CO]].

  M.~Papucci, A.~Strumia,
  JCAP {\bf 1003}, 014 (2010) 
  [arXiv:\hhref{0912.0742} [hep-ph]].

  G.~Zaharijas {\it et al.} [ for the Fermi-LAT Collaboration ],
   [arXiv:\hhref{1012.0588} [astro-ph.HE]] and talk by G.~Zaharijas at the \myurl{confluence.slac.stanford.edu/display/LSP/Fermi+symposium+2011}{FERMI Symposium 2011}.
      
\bibitem{Abramowski:2011hc}
  A.~Abramowski {\it et al.} [ H.E.S.S. Collaboration ],
  Phys.\ Rev.\ Lett.\  {\bf 106 } (2011)  161301.
  [arXiv:\hhref{1103.3266} [astro-ph.HE]].      
      
 \bibitem{collider}
J.~Goodman, M.~Ibe, A.~Rajaraman, W.~Shepherd, T.~M.~P.~Tait, H.-B.~Yu,
  Phys.\ Lett.\  {\bf B695 } (2011)  185-188 
  [arXiv:\hhref{1005.1286} [hep-ph]].
Y.~Bai, P.~J.~Fox, R.~Harnik,
  JHEP {\bf 1012 } (2010)  048,
  [arXiv:\hhref{1005.3797} [hep-ph]].
  J.~Goodman, M.~Ibe, A.~Rajaraman, W.~Shepherd, T.~M.~P.~Tait, H.-B.~Yu,
  Phys.\ Rev.\  {\bf D82 } (2010)  116010.
  [arXiv:\hhref{1008.1783} [hep-ph]].
     P.~J.~Fox, R.~Harnik, J.~Kopp, Y.~Tsai,
  [arXiv:\hhref{1103.0240} [hep-ph]].
  
  
  
\bibitem{Dolgov:1999wv}
  A.~D.~Dolgov, S.~H.~Hansen, S.~Pastor, D.~V.~Semikoz,
  Astropart.\ Phys.\  {\bf 14 } (2000)  79-90.
  [arXiv:\hhref{hep-ph/9910444} [hep-ph]].
  
\bibitem{kadanoffbaym}  
E.g.
  A.~Anisimov, W.~Buchmuller, M.~Drewes, S.~Mendizabal,
  Annals Phys.\  {\bf 326 } (2011)  1998-2038.
  [arXiv:\hhref{1012.5821} [hep-ph]].
Ê C.~Fidler, M.~Herranen, K.~Kainulainen and P.~M.~Rahkila,
Ê 
Ê[ arXiv:\hhref{1108.2309} [hep-ph]].
Ê 
Ê 
Ê 
Ê
Ê 
Ê M.~Beneke, B.~Garbrecht, C.~Fidler, M.~Herranen and P.~Schwaller,
Ê 
Ê Nucl.\ Phys. \ B {\bf 843} (2011) 177
Ê [arXiv:\hhref{1007.4783} [hep-ph]].
Ê 
Ê J.~S.~Gagnon and M.~Shaposhnikov,
Ê 
Ê 
Ê Phys.\ Rev. D 83 (2011) 065021
Ê [arXiv:\hhref{1012.1126} [hep-ph]].
Ê 
  T.~Konstandin, T.~Prokopec, M.~G.~Schmidt, M.~Seco,
  Nucl.\ Phys.\  {\bf B738 } (2006)  1-22.
  [\hhref{hep-ph/0505103}].
  V.~Cirigliano, C.~Lee, M.~J.~Ramsey-Musolf, S.~Tulin,
  Phys.\ Rev.\  {\bf D81 } (2010)  103503.
  [arXiv:\hhref{0912.3523} [hep-ph]].


\bibitem{Agrawal:2011ze}
  P.~Agrawal, S.~Blanchet, Z.~Chacko, C.~Kilic,
  [arXiv:\hhref{1109.3516} [hep-ph]].

\bibitem{MarchRussell:2009aq}
  J.~March-Russell, C.~McCabe, M.~McCullough,
  JHEP {\bf 1003 } (2010)  108.
  [arXiv:\hhref{0911.4489} [hep-ph]].

\bibitem{Baer:2011uz}
  H.~Baer, A.~Lessa, W.~Sreethawong,
  [arXiv:\hhref{1110.2491} [hep-ph]].
  
\bibitem{Buckley:2011ye}
  M.~R.~Buckley, S.~Profumo,
  [arXiv:\hhref{1109.2164v1} [hep-ph]].  
    
\end{thebibliography}
\end{document}